\documentclass[fleqn,usenatbib]{mnras}

\usepackage{newtxtext,newtxmath}

\usepackage{hyperref}

\usepackage[T1]{fontenc}
\usepackage{ae,aecompl}
\usepackage{soul}

\usepackage{graphicx}	
\usepackage{amsmath}	
\usepackage[usenames]{xcolor}
\usepackage{multirow}
\usepackage{booktabs}

\def\equationautorefname~#1\null{equation~(#1)}
\newcommand{\autorefp}[1]{%
  \begingroup%
  \def\equationautorefname~##1\null{equation~##1\null}%
  \autoref{#1}%
  \endgroup%
}
\usepackage{ulem}

\DeclareMathAlphabet{\mathpzc}{OT1}{pzc}{m}{it}
\definecolor{purple}{RGB}{160,32,240}

\newcommand{\nbodysix}{{\sc nbody6}}
\newcommand{\nbodytt}{{\sc nbody6tt}}
\newcommand{\Itid}{I_{\mathrm{tid}}}
\newcommand{\Mp}{M_{\rm p}}
\newcommand{\rh}{r_{\rm h}}
\newcommand{\rhi}{r_{\rm h,0}}
\newcommand{\rv}{r_{\rm v}}
\newcommand{\rhop}{\rho_{\rm p}}
\newcommand{\msqr}{\langle r^2\rangle}
\newcommand{\gm}{g_{\rm m}}
\newcommand{\tdynh}{t_{\rm dyn,h}}
\newcommand{\E}{\mathcal{E}}
\newcommand{\rt}{r_{\rm t}}
\newcommand{\timp}{t_{\mathrm{imp}}}
\newcommand{\dEimp}{\Delta E_{\mathrm{imp}}}
\newcommand{\rbound}{r_{\mathrm{h,bound}}}
\newcommand{\dr}{{\rm d}}

\title[Response of a star cluster to a tidal perturbation]{On the response of a star cluster to a tidal perturbation}

\author[Martinez-Medina et al.]{
Luis A. Martinez-Medina$^{1}$\thanks{Contact e-mail:\href{mailto:lamartinez@astro.unam.mx}{lamartinez@astro.unam.mx}}\href{https://orcid.org/0000-0002-5749-8255}{\includegraphics[scale=0.5]{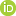}},
Mark Gieles$^{2,3}$\href{https://orcid.org/0000-0002-9716-1868}{\includegraphics[scale=0.5]{figs/orcid.png}},
Oleg Y. Gnedin$^{4}$\href{https://orcid.org/0000-0001-9852-9954}{\includegraphics[scale=0.5]{figs/orcid.png}}, and
Hui Li$^{5}$\thanks{NASA Hubble Fellow}\href{https://orcid.org/0000-0002-1253-2763}{\includegraphics[scale=0.5]{figs/orcid.png}}\\
$^1$Instituto de Astronom\'ia, Universidad Nacional Aut\'onoma de M\'exico, A.P. 70--264, 04510, M\'exico, CDMX, M\'exico.\\
$^2$ICREA, Pg. Llu\'{i}s Companys 23, E08010 Barcelona, Spain\\
$^3$Institut de Ci\`{e}ncies del Cosmos (ICCUB), Universitat de Barcelona (IEEC-UB), Mart\'{i} Franqu\`{e}s 1, E08028 Barcelona, Spain \\
$^{4}$Department of Astronomy, University of Michigan, Ann Arbor, MI 48109, USA\\
$^{5}$Department of Physics, Kavli Institute for Astrophysics and Space Research, MIT, Cambridge, MA 02139, USA
}

\date{Released \today}

\pubyear{2020}
\begin{document}
\label{firstpage}
\pagerange{\pageref{firstpage}--\pageref{lastpage}}
\maketitle

\begin{abstract}
We study the response of star clusters to individual tidal perturbations using controlled $N$-body simulations. We consider perturbations by a moving point mass and by a disc, and vary the duration of the perturbation as well as the cluster density profile. For fast perturbations (i.e. `shocks'), the cluster gains energy in agreement with theoretical predictions in the impulsive limit. For slow disc perturbations, the energy gain is lower, and this has previously been attributed to adiabatic damping. However, the energy gain due to slow perturbations by a point-mass is similar to, or larger than that due to fast shocks, which is not expected because adiabatic damping should be almost independent of the nature of the tides. We show that the geometric distortion of the cluster during slow perturbations is of comparable importance for the energy gain as adiabatic damping, and that the combined effect can qualitatively explain the results. The half-mass radius of the bound stars after a shock increases up to $\sim$7\% for low-concentration clusters, and decreases $\sim$3\% for the most concentrated ones. The fractional mass loss is a non-linear function of the energy gain, and depends on the nature of the tides and most strongly on the cluster density profile, making semi-analytic model predictions for cluster lifetimes extremely sensitive to the adopted density profile.
\end{abstract}

\begin{keywords}
galaxies: star clusters: general --- globular clusters: general --- open clusters and associations: general
\end{keywords}

\section{Introduction}
\label{sec:Intro}

Theories of globular cluster (GC) formation throughout cosmic time are becoming sophisticated enough \citep[e.g.,][]{muratov_gnedin10,li_gnedin14, kruijssen15, li_etal17, li_etal18, li_etal19, pfeffer_etal18, choksi_etal18, choksi_gnedin19} that accurate modeling of cluster dynamics in the early evolution has become more important. GCs are expected to form in environments where the gas density is high \citep{elmegreen_efremov97}, which leads to tidal perturbations by passing molecular gas clouds soon after GC formation. 

Much of our understanding of the response of a star cluster to a tidal perturbation is based on results of fast variation of the tidal field, i.e. `tidal shocks' \citep[][]{spitzer58, ostriker_etal72}. 
However, when the duration of the shock is longer than the crossing timescale of stars within the cluster -- which is short for dense clusters -- the response of the cluster is no longer described by the impulse approximation. During a slow perturbation, the stars conserve orbital actions, reducing the energy gain. The relevant parameter is the adiabatic parameter $x$, which is the ratio of the duration of the perturbation $\tau$ and the cluster dynamical time at the half-mass radius $\tdynh$. 
\citet{spitzer58} showed that for increasing $x$ the effect of the perturbation is adiabatically damped, leading to an exponential decrease of the energy gain with increasing $x$. In a series of important studies \citep{weinberg94a, weinberg94b, weinberg94c}, Weinberg showed that Spitzer's result underestimates the energy gain of slow perturbations because of resonances, leading to a power-law decrease of the energy gain with $x$. Understanding how the total energy gain of the cluster depends on $x$ makes it possible to generalize results from the impulse approximation to the adiabatic regime by introducing the concept of adiabatic correction.

Analytically, resonances in the adiabatic regime have been explored only for a star cluster passing through a one-dimensional slab that represents a galaxy disc \citep{weinberg94b, weinberg94c}.
Similarly, adiabatic corrections have been quantified by $N$-body simulations only for disc perturbations by \citet[][hereafter GO99]{gnedin_ostriker99} and for extended spherical perturbers by \citet{gnedin_etal99b}, who found that the cluster energy gain decreases as $x^{-3}$.
The resulting fitting expressions are often applied to other types of perturbers \citep{gnedin03a, prieto_gnedin08, pfeffer_etal18}, however the adiabatic corrections have not been quantified for other perturbers.
There is still a large unexplored region of the parameter space, such as the nature of the perturbers and the density profile of the cluster. In addition to disc-crossing, GCs also experience significant tidal forces from compact galactic structures such as galactic bulge or giant molecular clouds. These structures can be modeled as point mass (PM) since most encounters are distant ones such that their sizes are typically much smaller than the impact parameter to the cluster orbit. \citet{gieles06} showed that for a population of giant molecular clouds, most of the energy comes from penetrating encounters, for which the PM approximation does not hold. The results of such encounters can be understood from the combination of our 2 assumed models.
The distinction of different perturbers is meaningful in this context. The tidal forces of a disc perturbation are compressive in the $z$ direction, while half of the tidal forces of a passing PM are extensive, possibly leading to different forms of adiabatic corrections. 

The aim of this work is to go beyond the impulse approximation and explore the adiabatic regime of tidal perturbation through a set of controlled $N$-body experiments. We investigate the dependence of the energy gain on the nature of the perturber and properties of the cluster.
This paper is organized as follows. In \autoref{sec:nbody} we describe the setup of the $N$-body models and the parameterization of the tidal perturbations. \autoref{sec:results} contains our results of a study of the energy gain, mass loss, radius and density change. In \autoref{sec:discussion} we present a discussion on the intrinsic differences between a PM perturbation and a disc perturbation, and their impact on the adiabatic corrections. Finally, we summarize our results in \autoref{sec:summary}.

\section{Description of the $N$-body experiments}
\label{sec:nbody}

\subsection{$N$-body code}

We use a state-of-the-art direct $N$-body code \nbodytt, a modified version of the direct $N$-body integrator \nbodysix\ \citep{aarseth03} optimised for use with Graphics Processing Units \citep[GPUs,][]{nitadori_aarseth12}. It solves pairwise gravitational interactions between stars in the cluster. In order to apply single tidal perturbations to each cluster we use `Mode A' in \nbodytt, which applies tidal forces in the tidal approximation via user-defined tidal tensors as a function of time \citep*{renaud_etal11}. We adopt the canonical \cite{henon71} $N$-body units: $G = M_0 = -4E_0 = 1$, where $G$ is the gravitational constant, $M_0$ and $E_0$ are the total initial mass and energy of the cluster, respectively. For clusters in  virial equilibrium this results in an initial virial radius of $\rv \equiv -GM_0^2/(2W_0)=1$, where $W_0=2E_0$ is the initial gravitational energy.

\begin{table}
\centering
\begin{tabular}{ccc}
\toprule 
Shock $\dEimp/|E_0|$ & Shock duration $\tau$ & Cluster $W_0$ \\
\midrule
0.1  & 0.05, 0.1, 0.25, 0.5, 1, 2, 3, 4, 5 & 2, 4, 6, 8 \\
0.03 & 0.05, 0.1, 0.25, 0.5, 1, 2, 3, 4, 5 & 4 \\
0.01 & 0.05, 0.1, 0.25, 0.5, 1, 2, 3, 4, 5 & 4 \\
\midrule
1.0  & 0.05, 0.5 & 2, 4, 6, 8 \\
0.5  & 0.05, 0.5 & 2, 4, 6, 8 \\
0.1  & 0.05, 0.5 & 2, 4, 6, 8 \\
0.05 & 0.05, 0.5 & 2, 4, 6, 8 \\
\bottomrule
\end{tabular}
\caption{Parameter space explored in our $N$-body models. It encompasses a total of 172 simulations, 86 for each perturber: PM and disc.}
  \label{tab:parameters}
\end{table}

\begin{figure*}
\includegraphics[width=\textwidth]{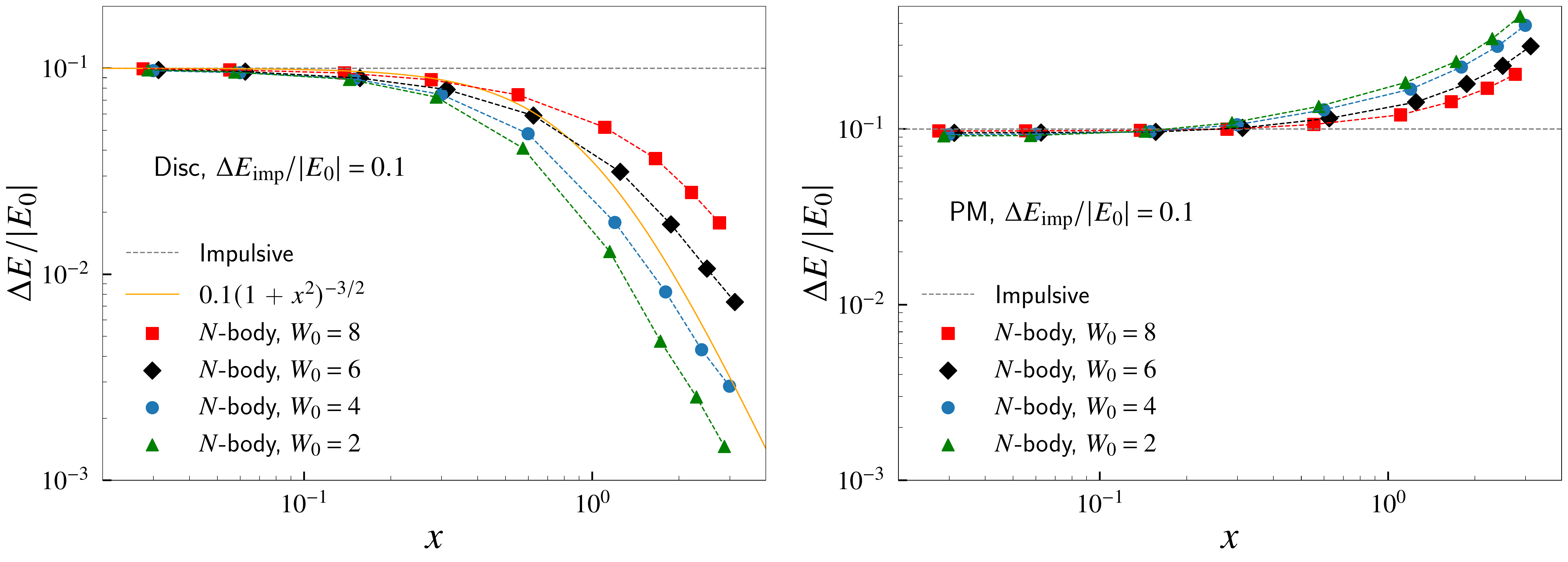}
\vspace{-5mm}
\caption{The fractional energy gain as a function of $x=\tau/\tdynh$, computed after the end of the disc (left) and PM tidal perturbation (right), with the expected $\dEimp/|E_0| = 0.1$  in the impulse approximation. Solid orange line in the left panel shows the adiabatic correction fit from GO99.} \label{fig:ShockA}
\end{figure*}

\subsection{Setup of the perturbers and tidal tensors}
\label{ssec:setup}
We perform experiments with two types of tidal perturbers: (i) a PM  with mass $\Mp$, and (ii)  an infinite disc with vertical density profile $\rho(Z) =\rhop\exp(-Z^2/H^2)$, where $H$ is the scale height and $\rhop$ is the mid-plane density. 

The duration of the perturbation is defined as $\tau \equiv b/V$ for the PM case, where $b$ is the impact parameter and distance of closest approach (which in the impulsive regime is the same as the impact parameter at infinity) and as $\tau \equiv H/V$ for the disc shock. To achieve a certain energy gain in the impulse approximation $\dEimp$, the parameter $\Mp$ in $N$-body units is found from equation~(9) of \citet{spitzer58}:
\begin{equation}
  \Mp = \frac{b^2V}{2G}\sqrt{\frac{3\dEimp}{\msqr}}.
  \label{eq:mp}
\end{equation}

For the disc shock we find that $\dEimp = (2/3)\msqr\gm^2/V^2$ \citep{ostriker_etal72}, where $\gm = 2\pi^{3/2}G\rhop H$ is the maximum acceleration, such that
\begin{equation}
  \rhop = \frac{V}{GH}\sqrt{\frac{3\dEimp}{8\pi^3\msqr}}.
  \label{eq:rhop}
\end{equation}
In all simulations we adopt $b=H=20$, but we note that in the tidal approximation these values are not important, because the strength of the tides are set by the density (see equations \ref{eq:Tij1} and \ref{eq:Tij2} below).

Each perturbation is applied to the cluster by computing the corresponding tidal tensor as a function of time. As the tidal tensor depends on the mass $\Mp$ for a PM shock, and on the density $\rhop$ for a disc shock, the inputs to compute the tensor are: the desired fractional energy gain in the impulsive approximation $\dEimp/|E_0|$, the duration $\tau$, the mean-square cluster radius $\msqr$ (which depends on the cluster model), and the type of the perturber (i.e. PM or disc). With this input, $\Mp$ and $\rhop$ are computed from equations (\ref{eq:mp}) and (\ref{eq:rhop}), respectively.

Then we compute the tidal tensor of a PM shock as:
\begin{equation}
    {\rm T}_{ij} \equiv -\frac{\partial^2\Phi}{\partial {X_i} \partial {X_j}}
    = \frac{G\Mp}{R^3}\left( \frac{3{X_i}{X_j}}{R^2} - \delta_{ij} \right),
\label{eq:Tij1}
\end{equation}
where $X_i$ are the coordinates of the PM with respect to the cluster, and $R^2 = \sum_i X_i^2$. The time dependent values of $X_i$ were computed for an orbit with $b$ and $V$ at infinity, meaning that as the orbit evolves the minimum distance (i.e. closest  approach) is smaller than $b$ and the relative velocity at closest approach is larger. For large $x$ this effect is important.

The tidal tensor of the disc is computed as:
\begin{equation}
    {\rm T}_{ij} = -4\pi G\rhop\exp\left(-\frac{Z^2}{H^2}\right)\, \delta_{i3}\delta_{3j}.
    \label{eq:Tij2}
\end{equation}
For the disc crossing we did not evolve the orbit but instead assumed a constant $V$. This was done because the infinity disc has a constant acceleration for $|Z|/H>>1$, which implies a strong, nonphysical correlation between the starting $Z$ and the ratio of the initial velocity and the maximum velocity.

In terms of the tidal tensor, the expected energy change in the impulse approximation can be written as
\begin{equation}
  \Delta E_{\rm imp} = \frac{1}{6}\, \Itid\, \msqr,
  \label{eq:itid}
\end{equation}
where the tidal heating parameter \citep{gnedin03a}
\begin{equation}
  \Itid \equiv \sum_{i,j} \left(\int T_{ij}\, \dr t\right)^2\ ,
\end{equation}
evaluates to
\begin{equation}
  \Itid = \frac{4\gm^2}{V^2} = \frac{16\pi^3 G^2 \rhop^2 H^2}{V^2}\ ,
\end{equation}
for the disc shock, and to 
\begin{equation}
  \Itid = \frac{8 G^2 \Mp^2}{b^4 V^2}\ ,
\end{equation}
for the PM shock.

We define the time of impact (or closest approach to the perturber) as $t=0$, such that most of the perturbation occurs between the times $-\tau$ and $\tau$. The values of $\dEimp/|E_0|$ and $\tau$ that we consider are given in \autoref{tab:parameters}.

It is worth mentioning that the temporal resolution for the sampling of the tensors scales with the duration of the encounter; it is 0.5 for the slowest perturbations, $\tau=5$, and improves up to 0.01 for the fastest shocks, $\tau=0.05$.

\subsection{Cluster setup}
\label{sec:clsetup}

We explore different density profiles for the cluster, given by \citet{king66} models with dimensionless central concentration parameter $W_0=[2,4,6,8]$. For these models, we find that the half-mass radius is $\rh = [0.849, 0.827, 0.804, 0.871]$ and ${\langle r^2\rangle} = [1.027, 1.184, 1.693, 3.242]$ in $N$-body units. Each cluster model was generated using {\sc limepy} \citep{gieles15}, and discretized into 100,000 equal-mass particles. 

For each cluster model and type of perturber we explore nine values of the shock duration, $\tau$, from fast to slow perturbations. The regime at which the shock occurs is given by the adiabatic parameter
\begin{equation}
  x \equiv \frac{\tau}{\tdynh},
  \label{eq:x}
\end{equation}
where $\tdynh$ is defined as in GO99:
\begin{equation}
  \tdynh\equiv \left(\frac{\pi^2\rh^3}{2GM} \right)^{1/2}.
  \label{eq:tdynh}
\end{equation}
We find $\tdynh \simeq [1.74, 1.67, 1.60, 1.81]$ in $N$-body  units, for $W_0=[2,4,6,8]$. Note that despite different density profiles, all cluster models we consider have similar values of $\tdynh$ as well as of a related quantity -- the average density at the half-mass radius.

Fast perturbations (i.e. shocks) have $x\ll1$, while slow perturbations have $x\gtrsim1$. We constructed models with different $x$ by changing the value of $\tau$. This combination of parameters gives us a total of 172 $N$-body models, summarized in \autoref{tab:parameters}. All simulations are run for $2\timp$+40 $N$-body times, where $\timp = {\rm int}(5\tau + 0.5)$ and the tensors are setup such that the maximum of the perturbation occurs at $\timp$. When plotting results as a function of time, we subtract $\timp$ from the $N$-body time, such that $t=0$ corresponds to the middle of the perturbation. By starting the perturbation at $t=-5\tau$, we miss the contribution from the tidal forces in the time interval $-\infty <t/\tau <-5$. By integrating the analytic expressions for the tidal acceleration  of \citet{spitzer58} over this time interval, we estimate that it contributes approximately 1\% to the total velocity change, and can therefore be safely ignored.

In order to test this estimation we have therefore repeated our slowest PM encounter ($x=3$) with a $W_0=4$ cluster, and added 40 time units to the ``pre-shock'' part of the original simulation. Hence, this new simulation runs for 40 + 10$\tau$ + 40 time units, i.e., now the encounter is symmetric. With this new simulation we found that the originally truncated part of the encounter accounts for 0.02\% of the total energy change. This result confirms that the truncated part of the encounters has a negligible effect.

\section{Results}
\label{sec:results}

In this section we describe the results of our {$N$-body} models and focus on the change in $E$, $M$ and $\rh$ as a result of the perturbation.

\subsection{Energy gain for different cluster models}
\label{sec:DE}

First, we quantify the total energy gain of all stars (bound and unbound) in a cluster model, $\Delta E$, due to a single PM or disc tidal perturbation.

\autoref{fig:ShockA} shows the fractional energy gain relative to the initial energy, $\Delta E/|E_0|$, computed at the end of the simulation for the 72 models with the expected $\dEimp/|E_0|=0.1$. For fast shocks ($x\lesssim0.1$) there is a good agreement with the expected values from the impulse approximation {for all models}. For the disc perturbations, the energy gain due to slow perturbations ($x\gtrsim 1$) is 1-2 orders of magnitudes lower than what is expected from the impulsive approximation. This reduced energy gain has been reported before and attributed to adiabatic damping.

Here it is worth mentioning that for the PM case we are using the values of $x$ computed at infinity; it is because as the orbit evolves, the relative velocity at closest approach is larger that the initial $V$, causing the actual $x$ value to change.

In the left panel of \autoref{fig:ShockA} we overplot the result of GO99 for disc perturbers and clusters with $W_0=4$
\begin{equation}
  \frac{\Delta E}{|E_0|} = \frac{\dEimp}{|E_0|} \left( 1+x^2 \right)^{-3/2}.
  \label{eq:ad_corr}
\end{equation}
The asymptotic behaviour $\Delta E \propto x^{-3}$ was derived by GO99 in the limit of slow perturbations, $x\gg1$. One power of $\tau$ comes from the expectation of the linear perturbation theory, and two powers of $\tau$ come from the normalization of the shock amplitude. Together this results in the expected scaling $\Delta E \propto \tau^{-3} \propto x^{-3}$. Equation~(\ref{eq:ad_corr}) matches the results of the GO99 $N$-body simulations for a cluster with $W_0=4$. 

We find that \autoref{eq:ad_corr} gives a fair description of the reduction in energy change for our disc $W_0=4$ models, however, it does not explain the results for the clusters with different $W_0$. The reduction is less for more concentrated clusters, which could be because these clusters have more stars with large periods which are in the impulsive regime. Alternatively, the larger envelopes of more concentrated models could lead to more geometrical distortion during slow perturbations. We discuss this further in \autorefp{sec:w0dep}.

\begin{figure}
\includegraphics[width=\columnwidth]{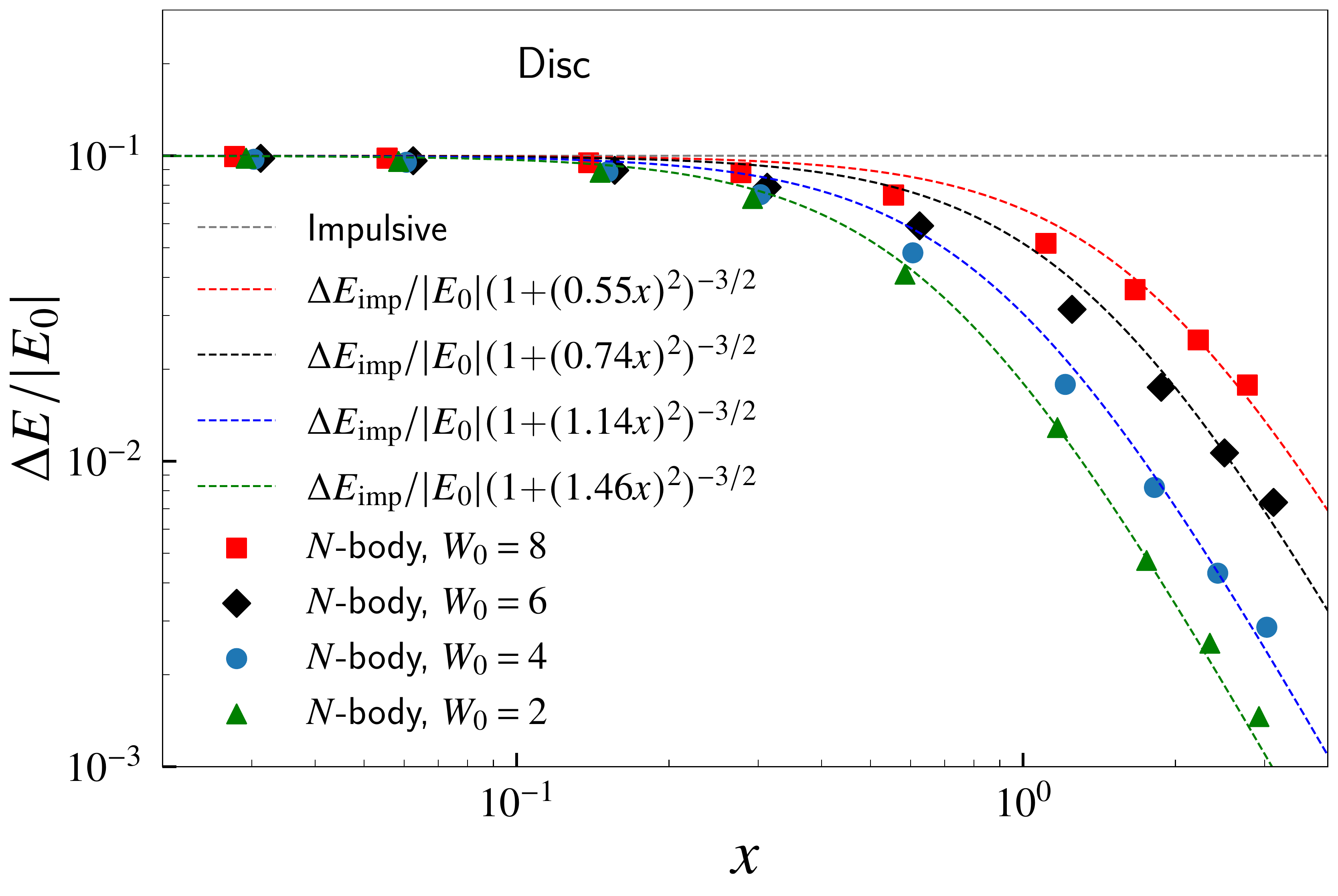}
\vspace{-5mm}
\caption{Energy gain of clusters of different $W_0$ for a disc tidal perturbation. Dashed lines are our fits given by \autoref{eq:ad_corr2}.}
\label{fig:Fit_3_2}
\end{figure}

However, we notice that by slightly changing this equation, we can still describe our disc perturbation models. To capture the dependence of our results on the density profile, we include an extra parameter $\epsilon$ that depends on $W_0$:
\begin{equation}
  \frac{\Delta E}{|E_0|} = \frac{\dEimp}{|E_0|} \left( 1+(\epsilon x)^2 \right)^{-3/2}.
  \label{eq:ad_corr2}
\end{equation}
\autoref{fig:Fit_3_2} shows the fits to our results using \autoref{eq:ad_corr2}, where the different values of $\epsilon$ illustrate the importance of including an additional parameter to take into account the density profile of the cluster. We find $\epsilon = [1.46, 1.14, 0.74, 0.55]$ for $W_0=[2,4,6,8]$.

The results for slow PM perturbations ($x\gtrsim 1$) are very different than those for the disc: for all cluster concentrations the energy gain is larger than what is expected from the impulsive approximation. It is also clear that none of the PM models can be described by \autoref{eq:ad_corr}.

Therefore we find that the difference with the impulsive prediction depends on the type of perturber and on the density profile of the cluster. We discuss possible causes of this behaviour in \autoref{sec:discussion}.

\subsection{Energy gain for different perturbation strengths}

Here we investigate a possible dependence of the cluster response on the perturbation amplitude. We apply perturbations of different strength to the $W_0=4$ model. For both perturbers, PM and disc, we take three values for the perturbation strength, $\dEimp/|E_0| = [0.01, 0.03, 0.1]$.

\begin{figure*}
\includegraphics[width=\textwidth]{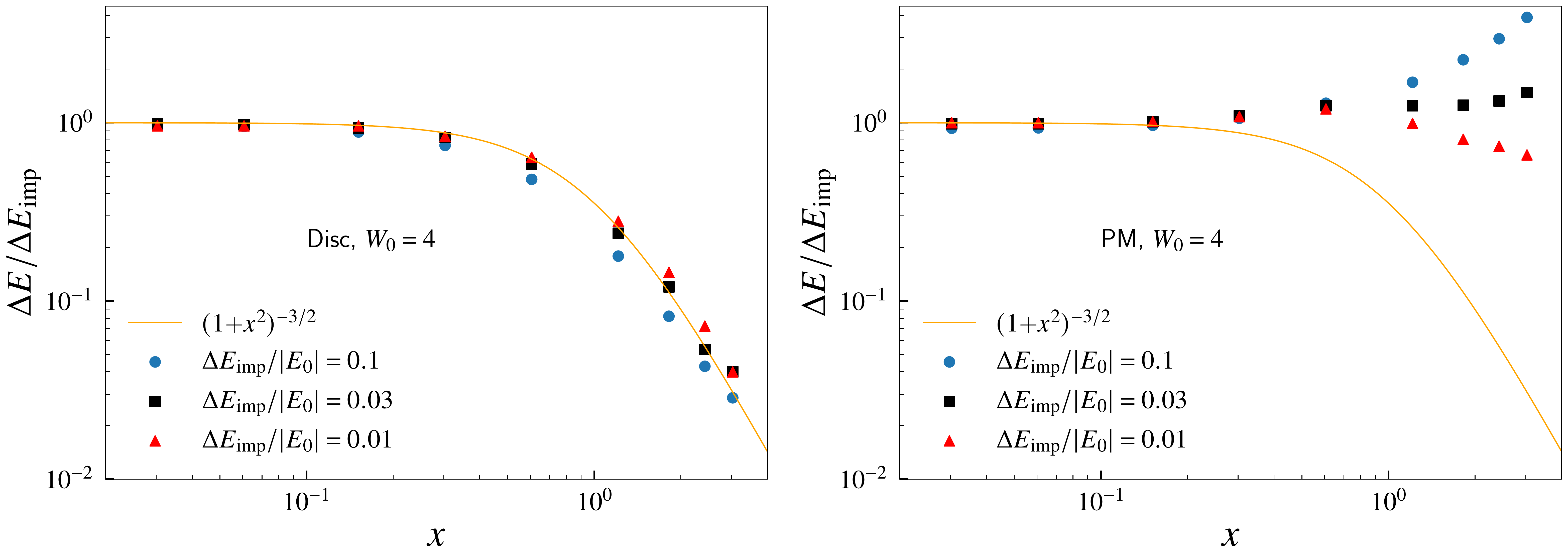}
\vspace{-5mm}
\caption{Fractional energy gain relative to the impulse approximation. Each panel shows three sets of $N$-body models that correspond to the shock strengths 0.1, 0.03 and 0.01, for cluster with $W_0=4$. The orange line shows the adiabatic correction fit from GO99.}
\label{fig:Strength}
\end{figure*}

To directly compare these cases, \autoref{fig:Strength} shows the ratio $\Delta E/\dEimp$ as a function of $x$. For the disc case, the cluster response is almost insensitive to the perturbation strength and is well described by \autoref{eq:ad_corr}. In contrast, for the PM case, the cluster response depends on the amplitude of the perturbation in the regime of $x>1$.

It is worth mentioning that GO99 tested \autoref{eq:ad_corr} in the range of disc shock strength from $\sim10^{-3}$ to $\sim1$. Here we reproduce that result for the $W_0=4$ models perturbed by a disc.

\subsection{Mass loss}
\label{ssec:dm}

The induced energy gain on the cluster causes some stars to escape the system, either by increasing their kinetic energy or reducing the binding energy. As a consequence, an imprint of energy to the cluster translates into a mass lost.

In our simulations $\Delta M$ comes from the particles that become unbound after the tidal perturbation. We define a particle as unbound if its specific energy $\E=0.5v^2+\phi$ is positive at the end of the simulation,\footnote{Note that we use a different symbol $\E$ to distinguish the specific energy of an individual particle from the total energy of the whole cluster $E$.} where $v$ is the velocity (measured in the non-rotating frame of the cluster's center of mass) and $\phi$ the specific potential due to the other stars (bound and unbound).

\begin{figure*}
\includegraphics[width=\textwidth]{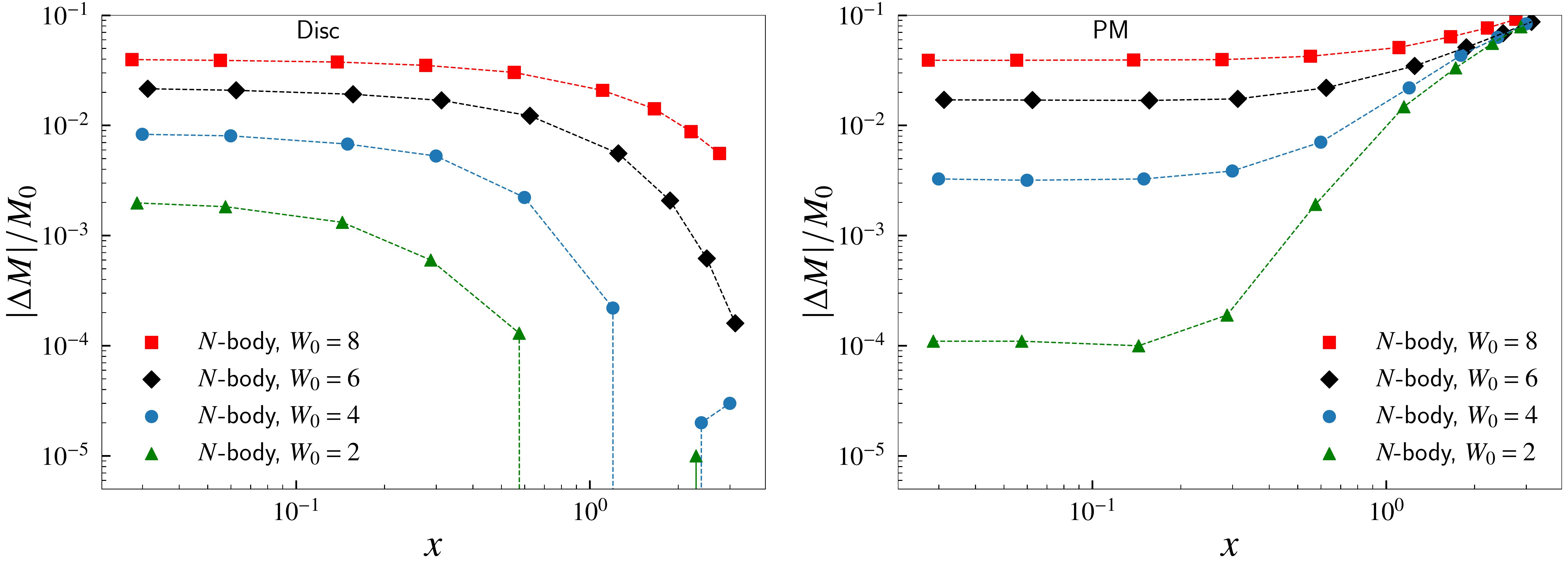}
\vspace{-5mm}
\caption{Fractional change in mass as a function of $x$, for a perturbation strength $\dEimp/|E_0|=0.1$}
\label{fig:DM_M}
\end{figure*}

\begin{figure}
\includegraphics[width=\columnwidth]{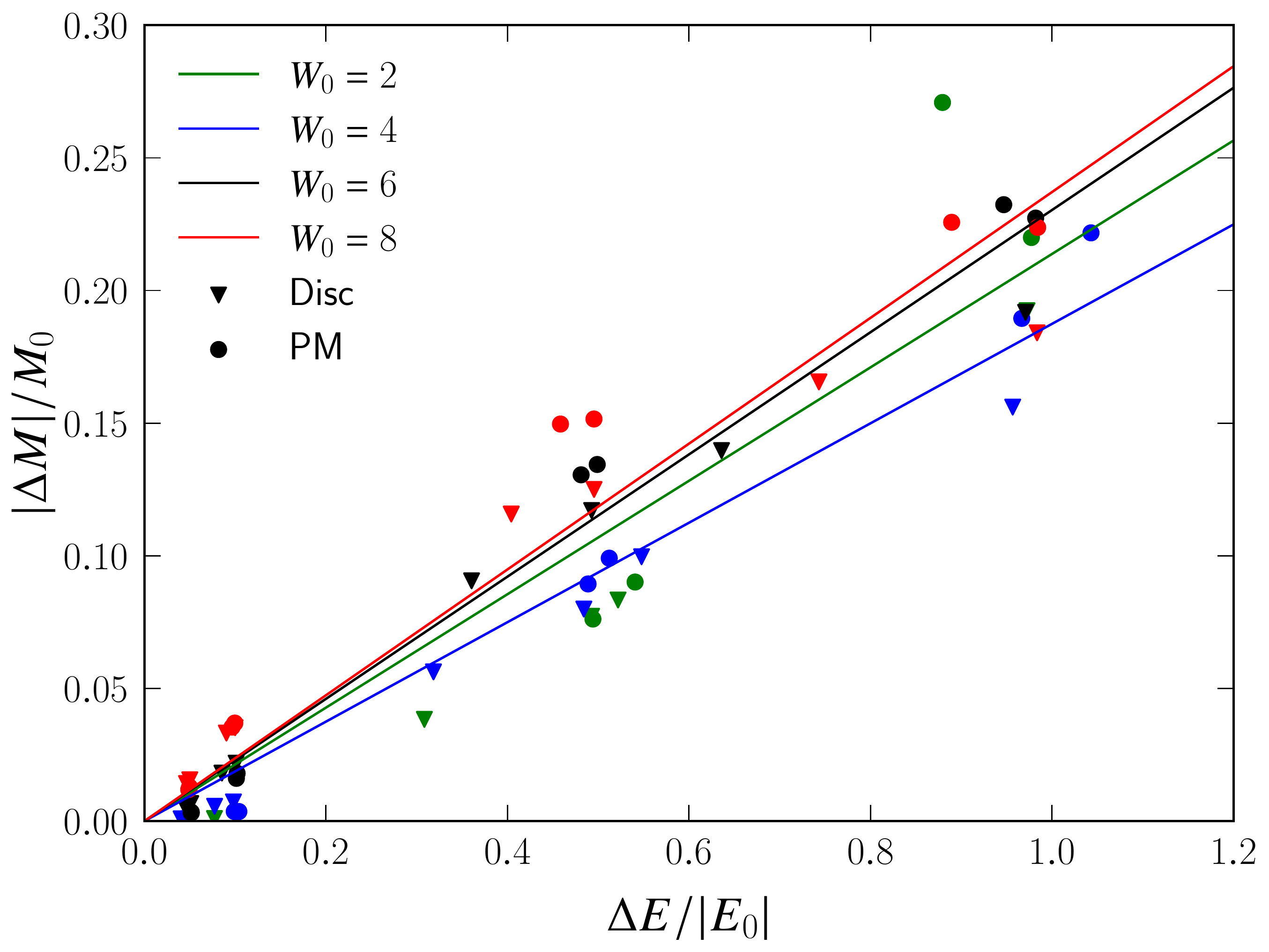}
\includegraphics[width=\columnwidth]{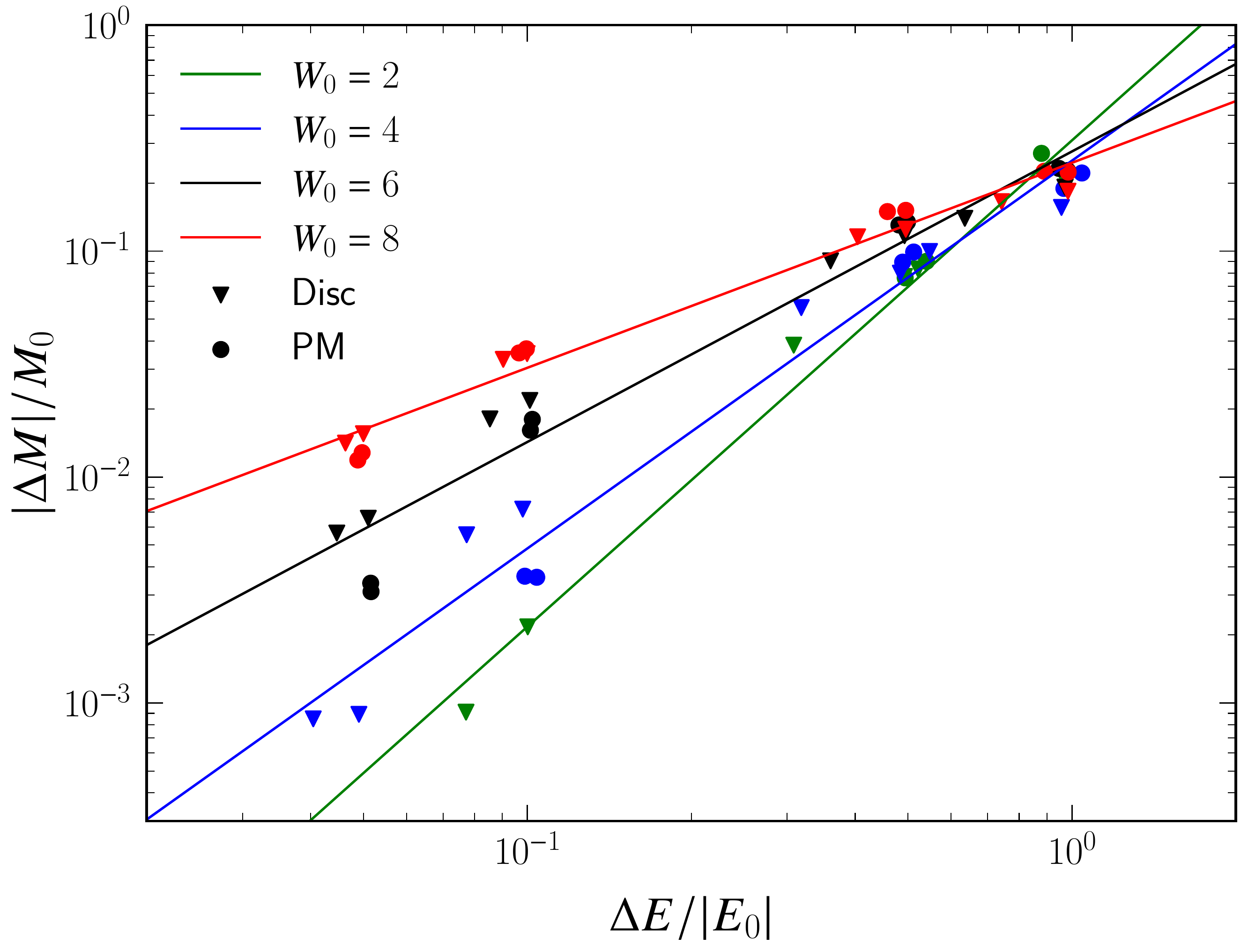}
\vspace{-4mm}
\caption{Fractional mass loss vs. energy change on a linear (top panel) and log-log scale (bottom panel). $N$-body results shown by symbols are differentiated by shock type and cluster density profile. Lines show linear regression fits, described in the text.}
\label{fig:DM_DE}
\end{figure}

In \autoref{fig:DM_M} we show the fractional mass loss as a function of $x$, for the disc (left) and PM (right) perturbations for $\dEimp/|E_0| = 0.1$. 
Notice that the value of the mass loss depends on the cluster concentration. The fractional mass loss is a factor of 20(400) higher for the disc(PM) shocks for $W_0 = 8$ clusters than for $W_0=2$ clusters.
For stronger shocks of $\dEimp/|E_0|\simeq0.5$ this difference reduces to a factor of $\sim2$, and for $\dEimp/|E_0|\simeq1$ there is no $W_0$ dependence (see \autoref{fig:DM_DE}).

To understand the $W_0$-dependence for weak shocks, we consider the fact that the maximum energy of particles in models with different $W_0$ is different: $\E_{\rm max} =-GM/\rt$ and $\rt$ for $W_0 = 8$ is about 3 times larger than for a $W_0=2$ model. One therefore expects that a $W_0=2$ model requires a larger $\Delta \E$ to unbind at least one particle. We need to keep in mind that the experiments were setup such that clusters have the same $\dEimp/|E_0|$, and because $\dEimp=\frac{1}{6}\Itid \msqr$ and $\msqr$  is approximately 3 times smaller for $W_0=2$ than for $W_0=8$, the value of $\Itid$ is 3 times larger for $W_0=2$ models. This means that at $\rt$ the expected energy gain ($\Delta \E_{\rm t} \propto \Itid\, \rt^2$) is 3 times larger for $W_0=8$ models, thereby unbinding more stars that are near $\rt$.

Meanwhile, in the adiabatic regime and for disc shocks, the cluster mass loss decreases as the perturbation becomes slower, following a trend similar to the energy gain shown in \autoref{fig:ShockA}.
For the PM perturbation $|\Delta M|/M_0$ behaves in the same way as for the disc in the impulsive regime, i.e., it is nearly constant, and its value strongly depends on $W_0$. However, for slow tidal perturbations, the cluster mass loss increases, opposite to what happens with the disc.

It is worth pointing out that although fast PM and disc  shocks converge to the same $\Delta E$ value (see \autoref{fig:ShockA}), they do not converge to the same $\Delta M$ value. \autoref{fig:DM_M} shows that, in the impulsive regime and for the same $\Delta E$, the cluster systematically losses more mass under a disc shock than under a PM shock; this difference is negligible for high concentration clusters ($W_0=8$), but becomes significant as the concentration decreases. For low concentrations ($W_0=2$) the cluster mass loss under a fast disc shock can be more than an order of magnitude higher than under a PM fast shock, for the same energy gain.

\begin{figure*}
\includegraphics[width=\textwidth]{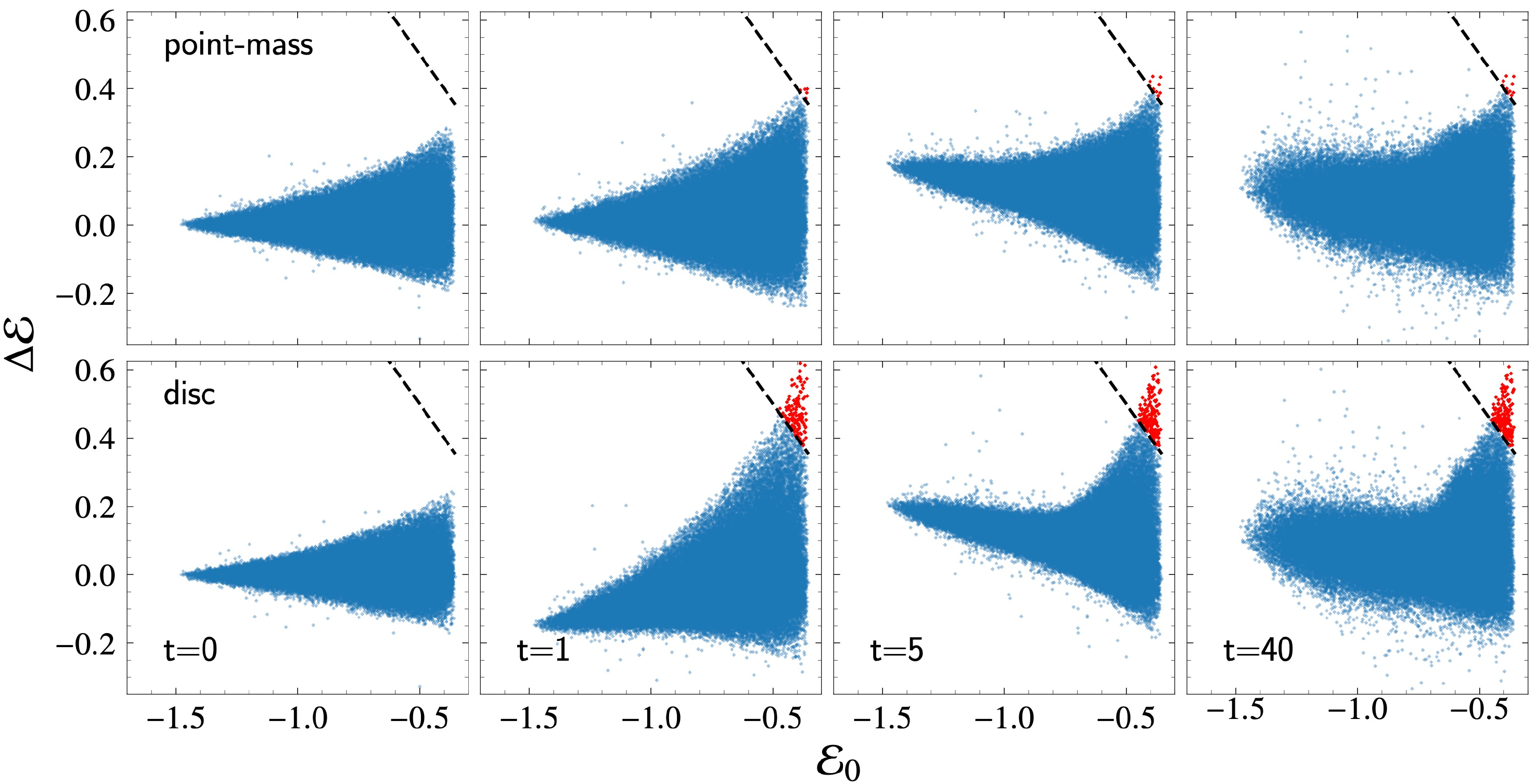}
\vspace{-5mm}
\caption{Energy gain of particles at different times (in $N$-body units) for an impulsive shock ($x=0.03$) due to a point-mass (top row) and a disc (bottom row). In both cases the cluster is described by a $W_0=2$ and the strength of the shock is $\dEimp/|E_0| = 0.1$. Stars that have become unbound are above the dashed line and are shown in red. More stars become unbound for a disc shock because all tidal forces are exerted along a single axis (see text for details).}
\label{fig:de_panel}
\end{figure*}

This discrepancy is likely the result of the different nature of the tidal forces: for the disc, the tidal forces are compressive along the $z$-axis, whereas for the PM case, half of the tides act along the $x$-axis (extensive) and the other half acts along the $z$-axis (compressive). This means that for the disc $\Delta v = \Delta v_z$, while for the PM $\Delta v \simeq \sqrt{\Delta v_x^2 + \Delta v_z^2}$ and $\langle \Delta v_x^2\rangle = \langle \Delta v_z^2\rangle=0.5\langle \Delta v^2\rangle$. Therefore the maximum energy gain of individual particles, $\Delta \E$, in the PM case is half of the maximum $\Delta \E$ in the disc case.
We illustrate this in \autoref{fig:de_panel} by showing the  $\Delta \E$ as a function of their initial energy $\E_0$ at various moments in time for both a PM  and a disc case. The first column shows $\Delta \E$ in the middle of the shock,  and the evolution proceeds to the right. The dashed lines indicate the boundary between bound and unbound. Soon after the shock ($t=1$), the loosely bound stars in the disc case have increased more in energy  than in the PM case, resulting in more unbound stars (red dots).

\autoref{fig:DM_DE} shows the fractional mass loss as a function of the energy gain for the 64 simulations described in the second part of \autoref{tab:parameters}, i.e., four values of the perturbation strength, two values of the duration, four values of $W_0$, and two perturber types. We show this relation first on a linear scale because this emphasizes the large values of mass loss which are of primary interest in studies of cluster evolution. Then we show it on a log-log scale to emphasize the weak shock regime.

There is large dispersion in mass loss at low $\Delta E/|E_0|$, which makes it difficult to find a simple relation between mass loss and energy gain. At large energy gains ($\Delta E/|E_0|>0.1$), the fractional mass loss is well approximated by a linear relation
\begin{equation}
  \frac{|\Delta M|}{M_0} \simeq f\, \frac{\Delta E}{|E_0|}, \hspace{8mm} f=0.217\,\pm 0.005,
  \label{eq:dm}
\end{equation}
with rms of the fit $\sigma = 0.022$. This result is consistent with the relation found by \cite{gieles06} ($f=0.22$) for \citet{plummer1911} models. It is worth mentioning that, although the error in the fit is relatively small, at low values of $\Delta E/|E_0|$ the scatter is not random: instead, the models with $W_0 \leq 4$ systematically fall bellow the linear fit, and those with $W_0 \geq 6$ fall on or above the fit.

We fit the linear mass--energy change relation individually for different $W_0$ models and find the proportionality factor $f$ to vary between 0.18 and 0.22 for the disc shocks, and between 0.20 and 0.26 for the PM shocks. Interestingly, the dependence on $W_0$ is not monotonic: the $W_0=4$ model has the smallest slope. Most concentrated clusters have the largest mass loss, as already seen in \autoref{fig:DM_M}.
The adiabatic parameter $x$ plays a smaller role, for the range of values we considered $(x<1)$.
These results also show the dependence on the perturber type. The mean values are $f = 0.195 \pm 0.006$ for disc, $0.233 \pm 0.008$ for PM, but they are still relatively close.

\autoref{fig:DM_DE} also indicates that the relation between mass and energy change can be alternatively described as non-linear. Indeed, we fit a power-law relation
\begin{equation}
  \frac{|\Delta M|}{M_0} =  f_\beta\, \left(\frac{\Delta E}{|E_0|}\right)^\beta, \hspace{6mm} f_\beta \approx 0.25
\end{equation}
and find the slope systematically decreasing from $\beta=2.01\pm0.13$ for $W_0=2$ to $\beta=0.88\pm0.06$ for $W_0=8$. The constant of proportionality of this relation is tightly constrained to be $f_\beta=0.23-0.26$. These relations are shown in the bottom panel of \autoref{fig:DM_DE}. We can see again that the more concentrated cluster models show larger relative mass loss and the less steep relation.

Despite the present scatter and non-linearity for weak shocks, the relation (\ref{eq:dm}) for strong shocks ($\Delta E/|E_0| > 0.1$) can be used in analytical or sub-grid models of cluster disruption, when the energy change from tidal shocks can be calculated. In addition to specifying the information about the shocks ($\Itid$), this requires assumptions about the cluster profile determining $\msqr$. To eliminate the latter assumption we can rephrase the mass loss of a cluster of a given density profile due to a perturbation with strength $\Itid$.

\begin{figure}
\includegraphics[width=\columnwidth]{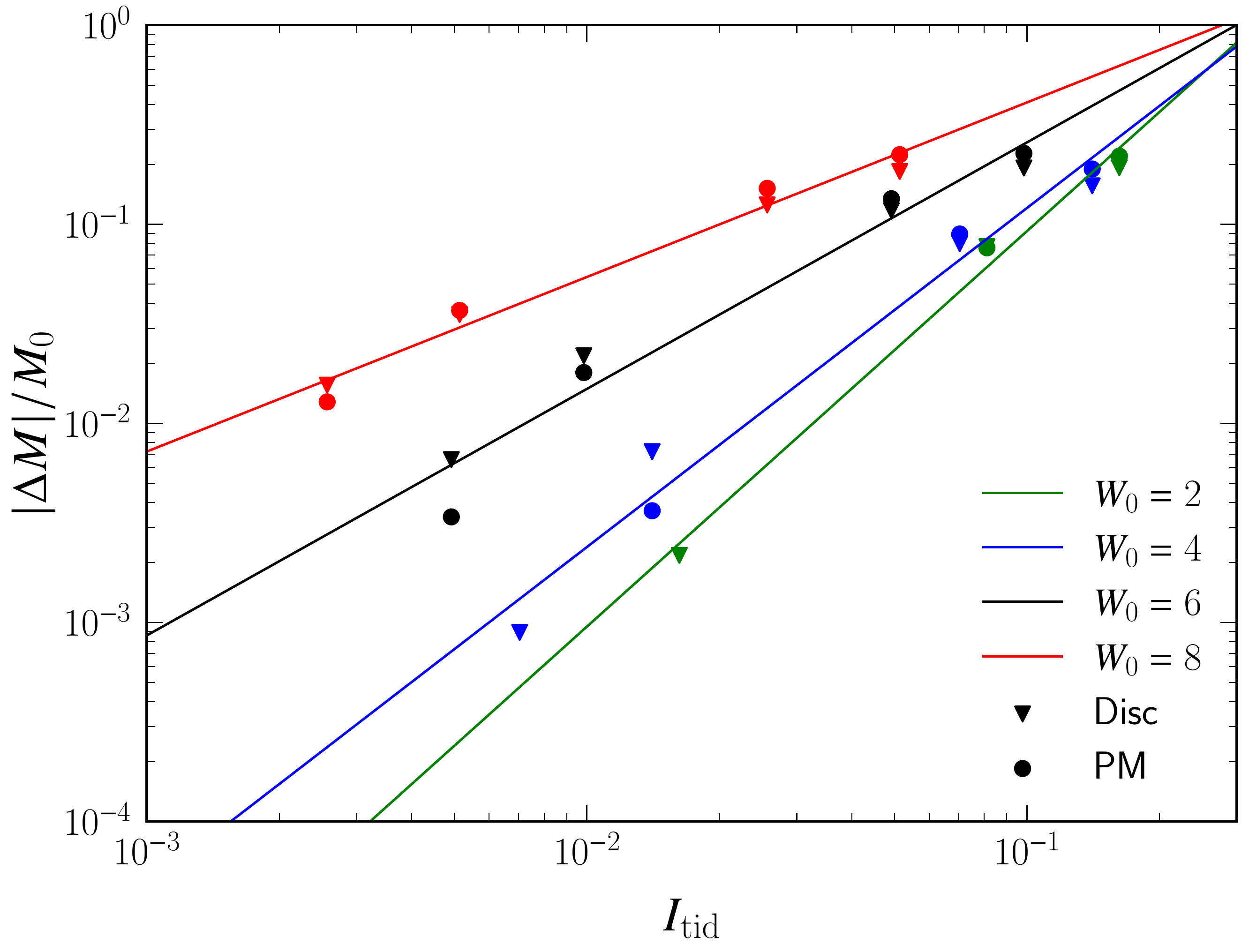}
\vspace{-4mm}
\caption{Fractional mass loss vs. tidal heating parameter $\Itid$, for the most impulsive runs ($x\approx0.03$). Symbols and colours are the same as in \autoref{fig:DM_DE}.}
\label{fig:DM_C}
\end{figure}

In \autoref{fig:DM_C} we plot the fractional mass loss as a function $\Itid$. The scatter is much larger than in \autoref{fig:DM_M}, because high(low) concentration clusters have moved to the left(right). This highlights the importance of the dependence of $\Delta E$ (and therefore $\Delta M$) on $\msqr$ of the cluster, which can vary by a factor of 3 for clusters with the same $\rv$ (and almost the same $\rh$).

We find that the relation between $\Delta E$ and $\Itid$ is almost exactly linear for a given cluster model $W_0$ (and therefore $\msqr$), as expected from analytical theory (\autorefp{eq:itid}): $\Delta E \propto \Itid^{0.98\pm0.01}$. However, the relation between $\Delta M$ and $\Itid$ is noticeably non-linear: 
\begin{equation}
  \frac{|\Delta M|}{M_0} \propto \Itid^\gamma
  \label{eq:dm_itid}
\end{equation}
with the power-law index decreasing systematically with $W_0$, from $\gamma=1.99\pm0.13$ for $W_0=2$ to $\gamma=0.88\pm0.06$ for $W_0=8$. These slopes are very close to those for the mass-energy relation, which means that the non-linearity of the $\Delta M(\Itid)$ relation is inherited from the $\Delta M(\Delta E)$ relation.

\begin{figure*}
\includegraphics[width=\textwidth]{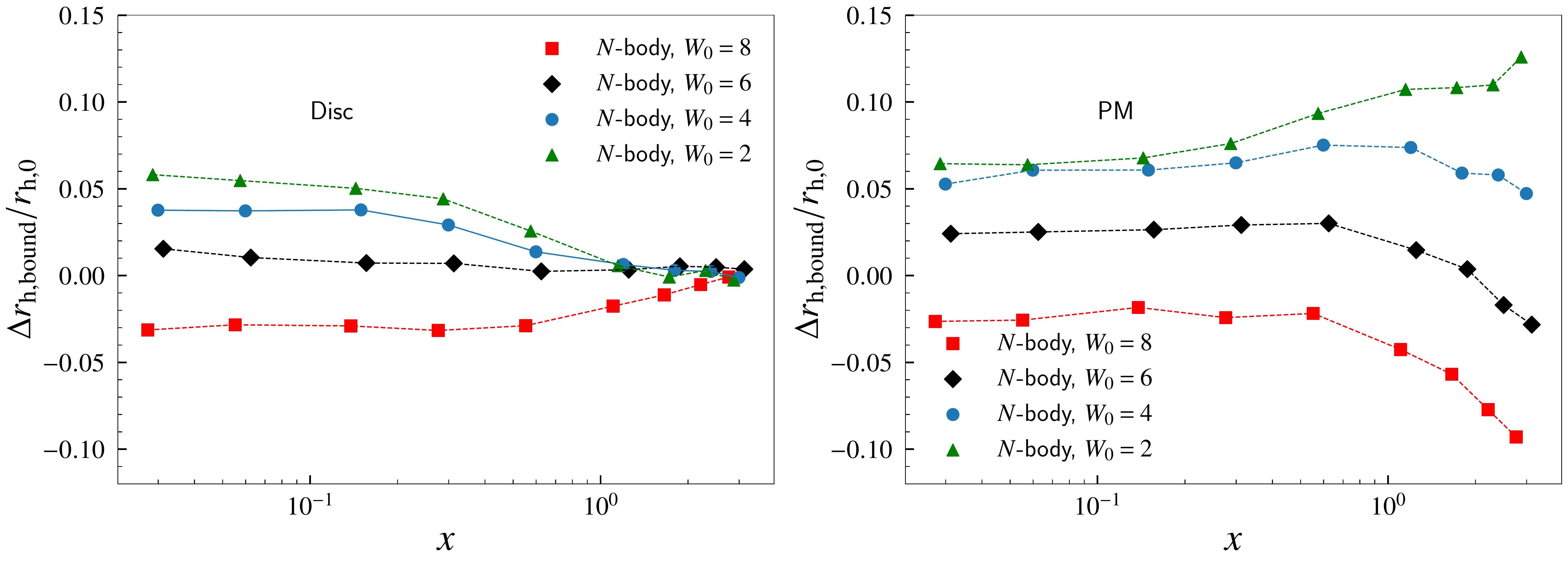}
\vspace{-5mm}
\caption{Fractional change in the half-mass radius of the bound stars as a function of the adiabatic parameter, for the perturbation strength $\dEimp/|E_0|=0.1$.}
\label{fig:DR_R}
\end{figure*}

\begin{figure*}
\includegraphics[width=\textwidth]{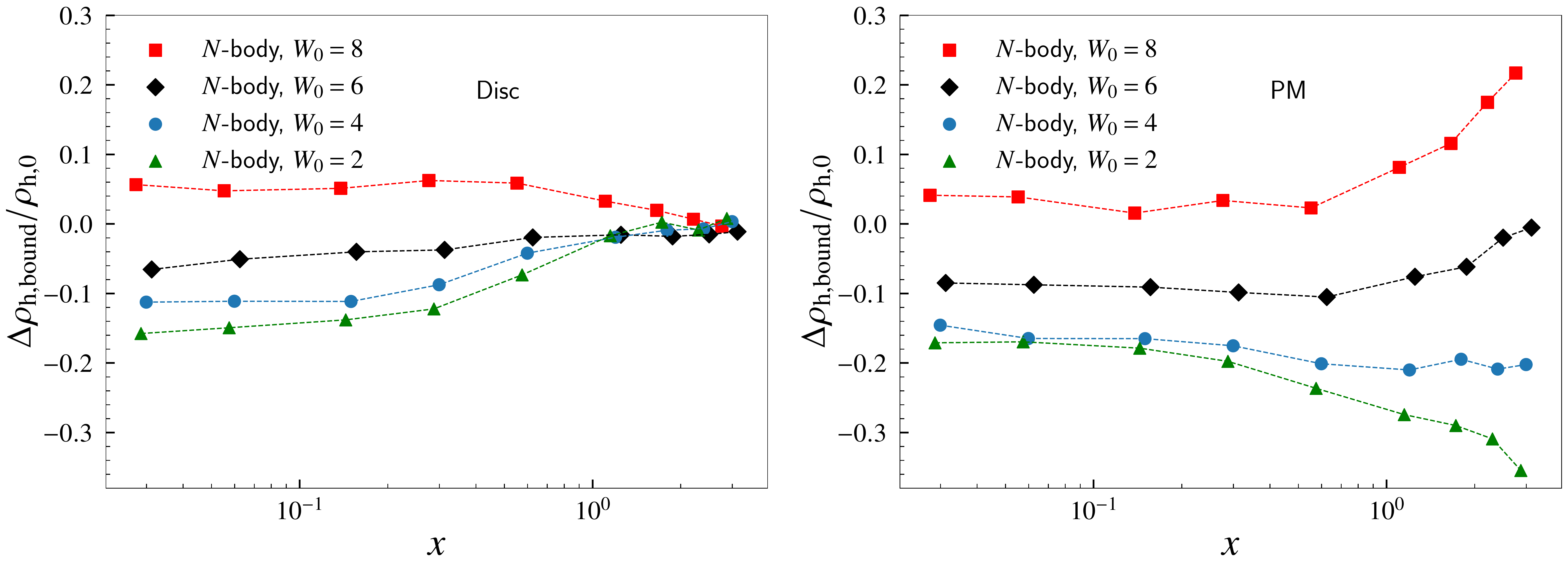}
\vspace{-5mm}
\caption{Fractional change in the average density of bound stars within the half-mass radius, $\rho_{\rm h,bound}$, as a function of adiabatic parameter, for a perturbation strength $\dEimp/|E_0|=0.1$.}
\label{fig:Drho}
\end{figure*}

\subsection{Changes in half-mass radius and density}

Besides the energy change and mass loss, $\rh$ of the cluster is also modified by the perturbation. Soon after the perturbations, the particles with the largest energy change will move to larger orbits thereby increasing $\rh$. However, the most energetic particles may be unbound and leave the cluster, such that the half-mass radius of the remaining bound stars, $\rbound$, could in fact be smaller than the initial half-mass radius, $\rhi$.

\autoref{fig:DR_R} shows the fractional change for the bound stars, $\Delta\rbound/\rhi$, as a function of $x$. First note that the change in $\rbound$ is larger for the less concentrated clusters regardless of the perturber. However, the bound system expands if the cluster is not too concentrated, while for our most concentrated model ($W_0=8$) it contracts -- $\Delta\rbound$ is negative. 
This is because for concentrated clusters, most of the added energy is absorbed by the distant stars which are already close to being unbound. Removing stars with near-zero energy results in a change of cluster mass at near-constant total energy, i.e. $E_{\rm tot}\propto M^2/\rh$ is approximately constant and therefore $\rh\propto M^2$ in the limit that the energy of the remaining bound stars is the same as the pre-shock energy. In less concentrated clusters, the mass that is lost also removes some negative binding energy, thereby increasing the energy of the remaining bound stars which leads to an increase of the radius \citep{gieles_renaud16}.

To have the full picture, we combine our measurements of mass and radius changes to compute the fractional change in the average density of bound stars within the half-mass radius, $\rho_{\rm h,bound}$. \autoref{fig:Drho} shows that after either a disc or PM perturbation, the bound system increases its density for the case of highest-concentration clusters, while for less concentrated clusters the density decreases. Whether the density increases or decreases defines whether the next shock is less or more disruptive, respectively.

As in the previous plots, \autoref{fig:Drho} also shows clear differences between a disc perturbation and a PM perturbation outside the impulsive regime. As the adiabatic parameter $x$ increases, the change in density of the high-concentration clusters grows in the PM case, while it decreases for the low-concentration clusters. On the other hand, in the disc case the changes in density converge to zero, as changes in both mass and radius diminish (Figs.~\ref{fig:DM_M} and \ref{fig:DR_R}).

Note that the changes in the cluster density imply a change in the value of the adiabatic parameter $x$. Because $\tdynh\propto \rho_h^{-1/2}$, the changes in $x$ and $\rho_h$ are related by $\Delta x/x = 0.5 \Delta \rho_h/\rho_h$. This means that the fractional changes in $\rho_h$ that we see in \autoref{fig:Drho} translate to changes in the value of $x$. We find that the adiabatic parameter oscillates for a few $N$-body times before returning close to its initial value. In the case of fast shocks, the amplitude of the oscillations can be up to 25\% of the initial value of $x$. In the case of $x\gtrsim1$ the variation is negligible. Therefore, the regime of the encounter (quantified by $x$) remains nearly the same as initial. 

\section{Discussion and interpretation}
\label{sec:discussion}

We have shown that a star cluster responds differently to a PM tidal perturbation than to a disc perturbation. Moreover, for each perturber and for a given amplitude of the applied shock (i.e. $\Itid$), the response depends also on the cluster concentration (i.e. $W_0$). Below we discuss possible reasons for this behaviour.

\subsection{The dependence of $\Delta E$ on cluster concentration for slow perturbations}
\label{sec:w0dep}

\autoref{fig:ShockA} shows that for slow perturbations with the same $\dEimp$, $\Delta E \ll \dEimp$ for discs, and $\Delta E > \dEimp$ for PMs. In previous work, the reduced energy gain for discs was ascribed to adiabatic damping. We first consider this effect to see whether it could explain the correlation of $\Delta E$ with $W_0$. High $W_0$ clusters may have a larger fraction of stars in the impulsive regime, reducing the effect of damping. To test this idea, we used the discrete realisations of King models with different $W_0$ from our initial conditions, and then analogous to \autoref{eq:tdynh} we calculated an estimate of the dynamical time for each particle as $t_{\rm dyn}(r) = [\pi^2 r^3/4GM(r)]^{1/2}$, where $r$ is the radial position of the particle and $M(r)$ is the enclosed mass at that radius. The fraction of stars with the local dynamical time exceeding the shock duration for our slowest perturbation, $t_{\rm dyn}(r) > \tau \approx 3\tdynh$ increases monotonically from 0.7\% for $W_0=2$ to 21\% for $W_0=8$. It is consistent with the trend of the parameter $\epsilon$ in \autoref{eq:ad_corr2} decreasing with $W_0$ from 1.46 to 0.55, indicating that the transition from impulsive to adiabatic occurs at higher $x$ for larger $W_0$. However, this does not explain fully the qualitative difference between the results for the disc and PM cases shown in \autoref{fig:ShockA}.

We therefore also look at the effect of geometrical distortions during the perturbations. For this estimate, we first consider the disc shock because of its simple one-dimensional tidal action, and the two extreme $W_0$ clusters in the suite: $W_0=2$ and $W_0=8$. We denote their respective quantities with subscript $2$ and $8$. Consider a slow perturbation with a given $\dEimp = \frac{1}{6}\, \Itid\, \msqr$. Because $\msqr_8\simeq 3\msqr_2$ (\autoref{sec:clsetup}) and $\dEimp$ is the same for both clusters, we have  $I_2 \simeq 3 I_8$. The velocity change for a particle at $r\simeq\rt$ is $\Delta v \propto \sqrt{\Itid}\rt$, and because ${\rt}_{,8} \simeq 3{\rt}_{,2}$ (\autoref{sec:clsetup}), we find that $\Delta v_8/\Delta v_2 \simeq \sqrt{3}$.  The fractional compression along $z$ direction in the first half of the perturbation is $F \propto  \Delta v/\rt$, and therefore the ratio $F_2/F_8 \propto \sqrt{I_2/I_8} = \sqrt{3}$. This means that, somewhat counter-intuitively, less concentrated clusters (low $W_0$) are more compressed during the perturbation than concentrated clusters. This could also explain why the final $\Delta E$ of low-concentration cluster is lower, because their $\langle z^2\rangle$ are reduced more relative to the initial ones during the second half of the perturbation.

The PM results can be explained with these simple geometrical arguments: half of the tidal forces are compressive, hence the same arguments of the disc apply. However, the other half are extensive, and this should lead to a larger fractional extension for low$-W_0$ clusters, resulting in a larger $\Delta E$ for low concentration clusters. If we apply symmetry arguments, we expect the effect of compression on $\Delta E$ to cancel the effect of the extension, predicting that the energy gain for slow PM perturbations is the same as for fast shocks. 

\begin{figure*}
\includegraphics[width=\textwidth]{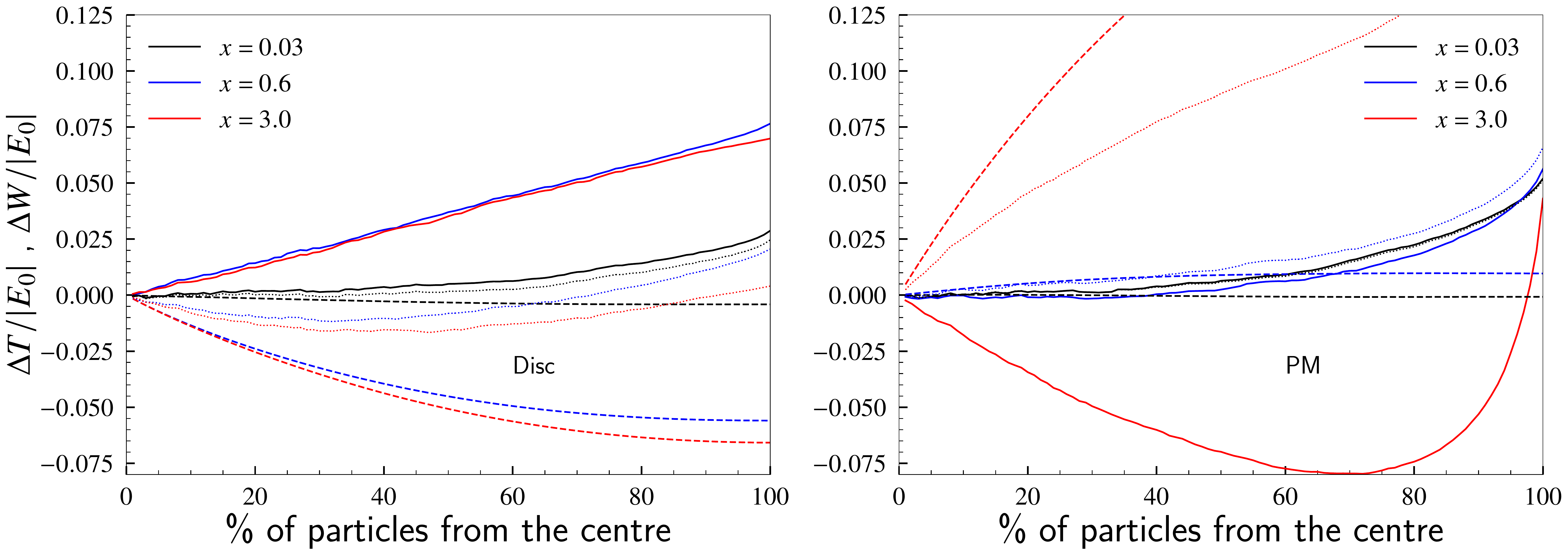}
\caption{Cumulative gains in kinetic energy $\Delta T/|E_0|$ (solid lines), potential energy $\Delta W/|E_0|$ (dashed lines), and total energy $\Delta E/|E_0|$ (dotted lines), computed at the middle of the encounter for a disc (left) and PM perturbation (right), on a $W_0=4$ cluster. The colours indicate different values of the adiabatic parameter.}
\label{fig:DT_DW}
\end{figure*}

\subsection{On the different response of a star cluster to a disc vs. PM perturbation}
\label{sec:tides}

To explore in more detail the cause of the differences between a disc and a PM tidal perturbation we split the total energy gain into the kinetic ($K$) and potential ($W$) energy parts. For this exercise we look at the initial configuration and at the middle of the shock. Each of the clusters is binned in radial shells such that each shell contains the same number of particles, in this case we choose 1\% of the total of particles per shell. For each shell, we compute the cumulative kinetic and potential energy of the particles contained within the shell. Then we subtract the initial $K$ and $W$ from the values at the middle of the shock. With this procedure we obtain the cumulative $\Delta K$ and $\Delta W$ across the cluster.

\autoref{fig:DT_DW} shows $\Delta T/|E_0|$ and $\Delta W/|E_0|$ as a function of the percentage of particles from the cluster centre. Note that the evolution of kinetic and potential energies depends strongly on the nature of the perturber.
The gain in kinetic energy $\Delta T$ for the PM perturbation decreases as the duration of the perturbation increases, but has the opposite behavior for the disc.
The gain in potential energy $\Delta W$ for the PM perturbation increases ($W$ becomes less negative) as the perturbation enters into the adiabatic regime, while for the disc $W$ becomes more negative.

The reason for the opposing behaviours in both $\Delta T$ and $\Delta W$ lies in the different directions of the tides for the two types of perturber. The PM tides have equal extensive and compressive components \citep{spitzer58}, while they are fully compressive for a disc case. As shown in \autoref{fig:ShockA}, the nature of the tide is not important for the energy gain in the impulsive regime, but it becomes important when the duration of the perturbation increases.

\autoref{fig:DT_DW} shows that in the adiabatic regime, as $x$ gets closer to 1, the extensive tide due to the PM has enough time to pull the stars apart, decreasing the velocity dispersion ($\Delta T<0$) and reducing the binding energy ($\Delta W>0$). On the other hand, the compressive tide due to the disc pushes the stars inwards, increasing the velocity dispersion ($\Delta T>0$) and making the cluster more bound ($\Delta W<0$).

In the formulation of tidal heating by \citet{weinberg94b}, a star's energy changes when it passes through a orbital resonance with the external gravitational potential. The number and distribution of resonances is particularly important for a slow perturbation where the gradually evolving stellar orbit may go through multiple resonances. As described by \citet{murali_weinberg97}, a spherical potential has a different (discrete) spectrum of resonances than the disc potential (continuous spectrum). This may be another reason for the different cluster response to the two perturbers.

\subsection{The effect of the orbit}

Another effect that could be important is that in the PM perturbation the impact parameter and relative velocity are time dependent, while for the disc we fix it (see Section~\ref{ssec:setup}). Here we estimate how much the PM orbit deviates from a straight line. We can do this by realising that for a straight-line orbit the kinetic energy of the orbit is much larger than its gravitational energy, i.e. $2G\Mp/(bV^2)\ll1$. With the expression for $\Mp$ from equation~(\ref{eq:mp}) and the definition of $\tau$ and $x$ we then find that the orbit is approximately a straight line when

\begin{equation}
    x\lesssim 1.6\sqrt{\frac{0.1}{\dEimp/|E_0|}}.
\end{equation}

Here we used the $\msqr$ values for $W_0=6$. 
From this we see that the 3 slowest encounters in Fig.~\ref{fig:ShockA} are affected by this, but the deviation from a straight-line orbit can not explain the difference between the behaviour of slow perturbations by a disc and a PM. The dependence on $\dEimp$ may explain why the weaker (slow) PM perturbations in Fig.~\ref{fig:Strength} show a larger $\Delta E$, their orbits are more gravitationally focused and have a smaller impact parameter ($p$) which leads to a larger energy gain. The maximum velocity ($V_{\rm max}$) is higher than $V$, but this does not compensate the smaller $p$, because the energy gain depends on $p^{-4}V_{\rm max}^{-2}$.

\subsection{The shape of the cluster}
In Sec. \ref{sec:DE} we found that not all our $N$-body models can be described by the analytic result for the energy gain of GO99 (\autorefp{eq:ad_corr}). The result of GO99 is a fitting function with the theoretically expected power-law slope at large $x$ ($x^{-3}$) and GO99 did not vary their cluster density profile to study its effect. Because of this, in this section we contemplate the possibility that the size and shape of the cluster are changing during the encounter, which could explain the disagreement between our simulations and \autoref{eq:ad_corr}, and between the disc and PM perturbation.

To quantify the deformation of the cluster due to the tidal perturbation we need to measure its shape. Following \citet{zemp_etal11}, we calculate the shape tensor, defined by the matrix components:
\begin{equation}
  \label{eq:shape}
  S_{jk} = \frac{\sum_i m_i \, x_{ij} \, x_{ik}}{\sum_i m_i},
\end{equation}
where index $i$ runs over $i=1,...,N$ stars, and indices $j$ and $k$ denote the three Cartesian coordinates for star $i$. The principal axes of the cluster can be interpreted as the eigenvectors of $S_{jk}$. The lengths of the primary semi-axes $a\geq b\geq c$ correspond to the square root of the eigenvalues. We compute the matrix $S_{jk}$ for all stars in our $W_0=4$ models under the fast and slow perturbations. 

\begin{figure*}
\includegraphics[width=\textwidth]{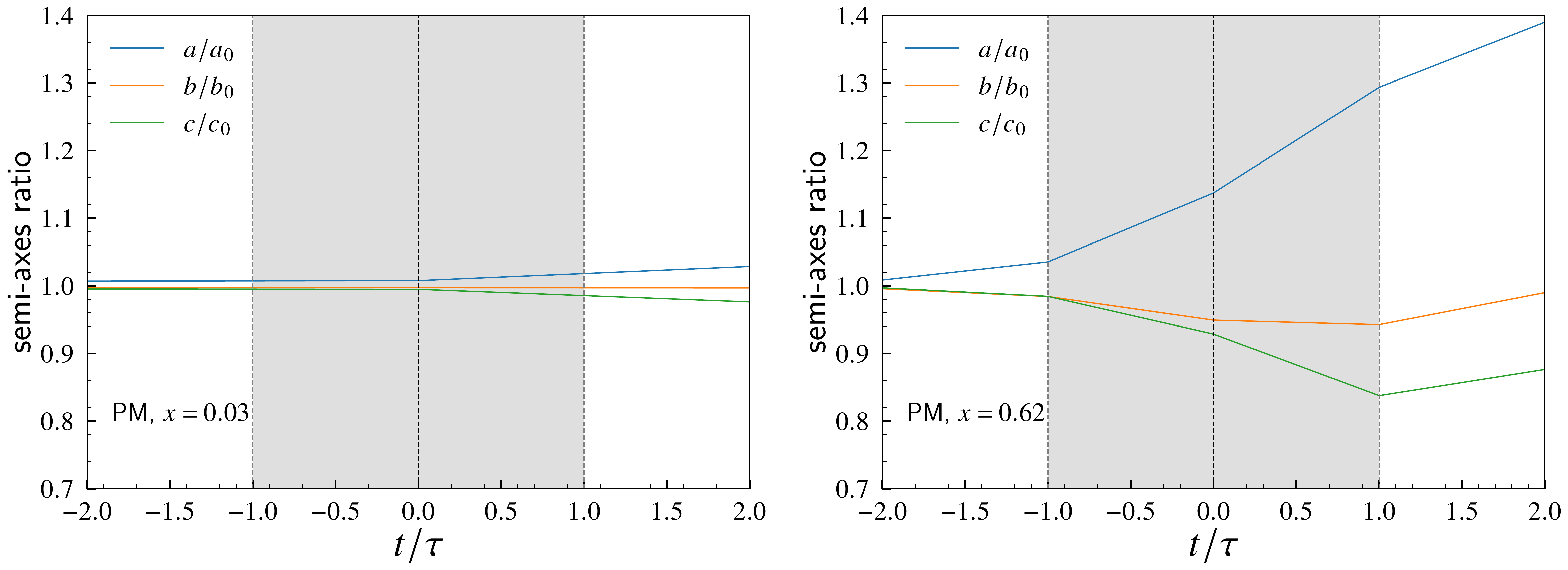}
\vspace{-4mm}
\caption{Temporal evolution of the cluster semi-axis lengths for a fast PM shock ($x=0.03$) and a slow PM tidal perturbation ($W_0=4, x=0.62$), both with $\dEimp/|E_0|=0.1$. The dashed black line indicates the time of maximum approach, $t=0$, while the grey shaded area covers the time interval [$-\tau,+\tau$], where $\tau$ is the duration of the perturbation.}
\label{fig:shapePM}
\end{figure*}

\autoref{fig:shapePM} shows the evolution of the semi-axes of the cluster as a function of time for the PM case and $\dEimp/|E_0|=0.1$. The grey shaded area indicates the time interval $\pm\tau$ where most of the perturbation occurs. We see that as the perturbation goes on, the cluster is slightly stretched along one of its axis, and compressed along the other two. Notice that the magnitude of the changes in the cluster shape are significantly larger for slow perturbations than for the fast ones.

\begin{figure*}
\includegraphics[width=\textwidth]{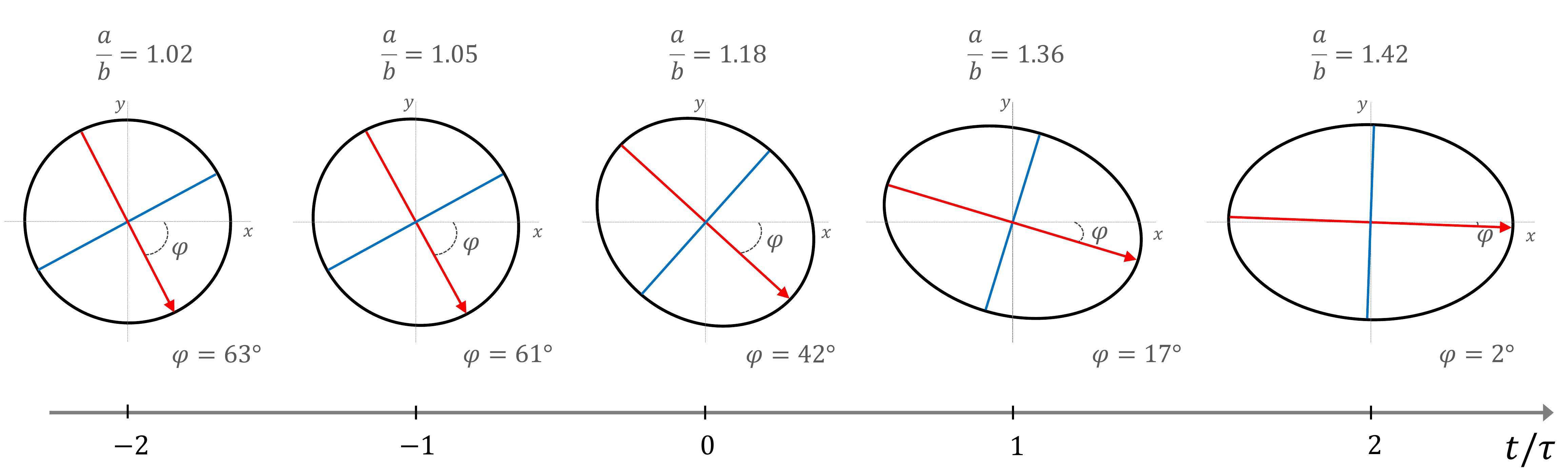}
\vspace{-4mm}
\caption{Temporal evolution of the cluster semi-axis direction for a PM tidal perturbation ($W_0=4, x=0.62$), with $\dEimp/|E_0|=0.1$. The cluster is at the origin while the PM moves along the $y$ direction, which imprints a rotation to the cluster. We show the same time interval as in \autoref{fig:shapePM}, with $t=0$ being the moment of closest approach. We indicate the angle between the semi-major axis (red arrow) and the $x$ axis, as well as the ratio of the semi-lengths $a/b$.}
\label{fig:shape}
\end{figure*}

Recall that initially the cluster is located at the origin of the coordinate system while the PM starts moving in the $y$ direction (with a closest approach of $b$ for fast shocks, see \autorefp{eq:mp}), hence, the long and middle semi-axes lie in the $x-y$ plane and their directions rotate during the flyby. \autoref{fig:shape} illustrates the changing orientation of the semi-axes as a function of time for one model. For this plot we computed the eigenvectors of the shape tensor of the cluster for snapshots before, during, and after the moment of closest approach. In the beginning of the perturbation the cluster is slightly stretched but still maintains a nearly spherical shape; then, at the moment of closest approach, the elongation of the cluster is noticeable, and keeps increasing after the PM has passed. The degree of deformation of the cluster is quantified with the axis ratio $a/b$, which increases with time. Note how the semi-major axis rotates due to the gravitational attraction exerted on the cluster by the moving PM.

\begin{figure*}
\includegraphics[width=\textwidth]{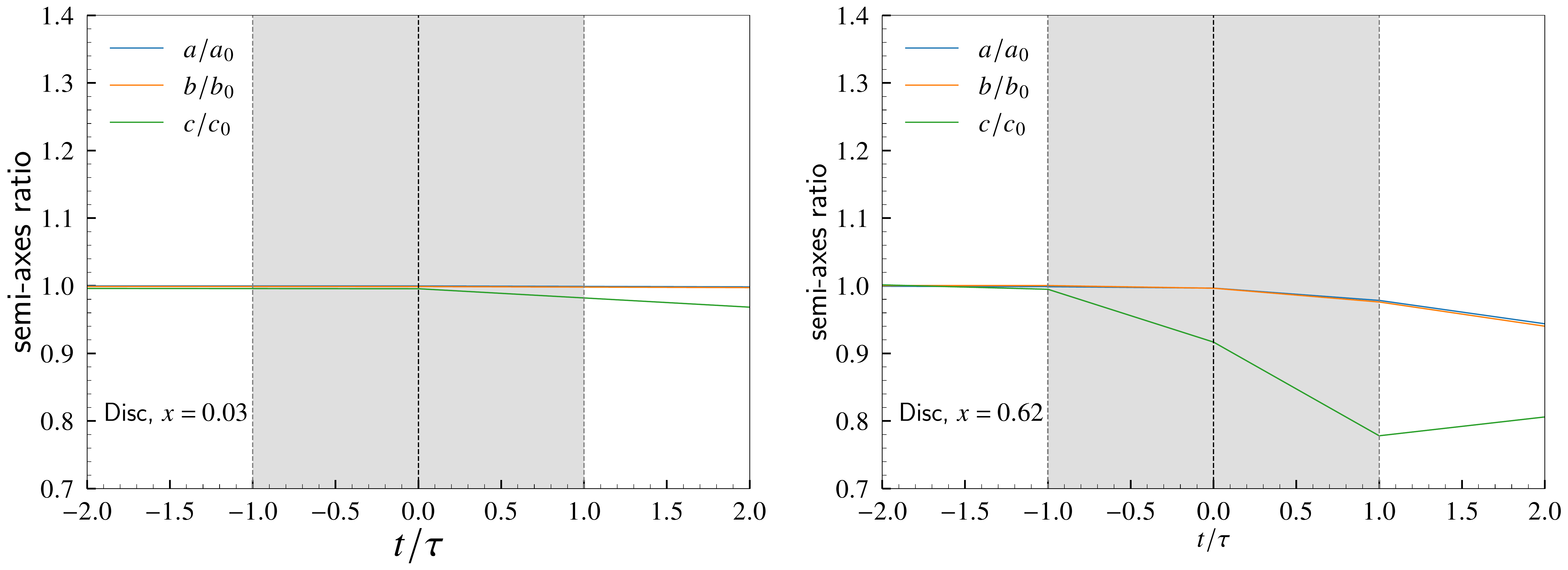}
\vspace{-4mm}
\caption{Temporal evolution of the cluster semi-axes length for a fast disc shock ($x=0.03$) and a slow disc tidal perturbation ($x=0.62$), both with $\dEimp/|E_0|=0.1$. The dashed black line indicates the time of maximum approach, $t=0$, while the grey shaded area covers the time interval [$-\tau,+\tau$], where $\tau$ is the duration of the perturbation.}
\label{fig:shapedisc}
\end{figure*}

On the other hand, in the case of a disc perturbation the cluster moves in the $z$ direction and crosses the disc located in the $x-y$ plane. \autoref{fig:shapedisc} shows that during the perturbation the $x-$axis and $y-$axis of the cluster remain unchanged, while it is compressed along its $z-$axis. Again, the change in the size of the cluster is significant for slow perturbations, but negligible for the fast ones.

The main difference in cluster response to the PM and disc perturbations is in the sign of the size change. The remaining bound cluster grows in size when the perturber is a PM, while it is compressed when the perturber is a disc.

\subsection{Expected change in cluster shape}

We note that the fractional change of the semi-axis length depends not only on the perturbation strength, but also on its duration.

In this section we characterize the expected change of cluster shape as a function of $x$ and $\dEimp/|E_0|$. In order to do this we use the $W_0=4$ subset from the first row of simulations in \autoref{tab:parameters} (perturbation strength 0.1), together with the second and third rows, also with $W_0=4$ with perturbation strengths 0.03 and 0.01. Again, we compute the length of the primary semi-axes of the cluster as the square roots of the eigenvalues of the shape tensor (\autorefp{eq:shape}) at a time $\tau$ after the impact.

\begin{figure}
\includegraphics[width=\columnwidth]{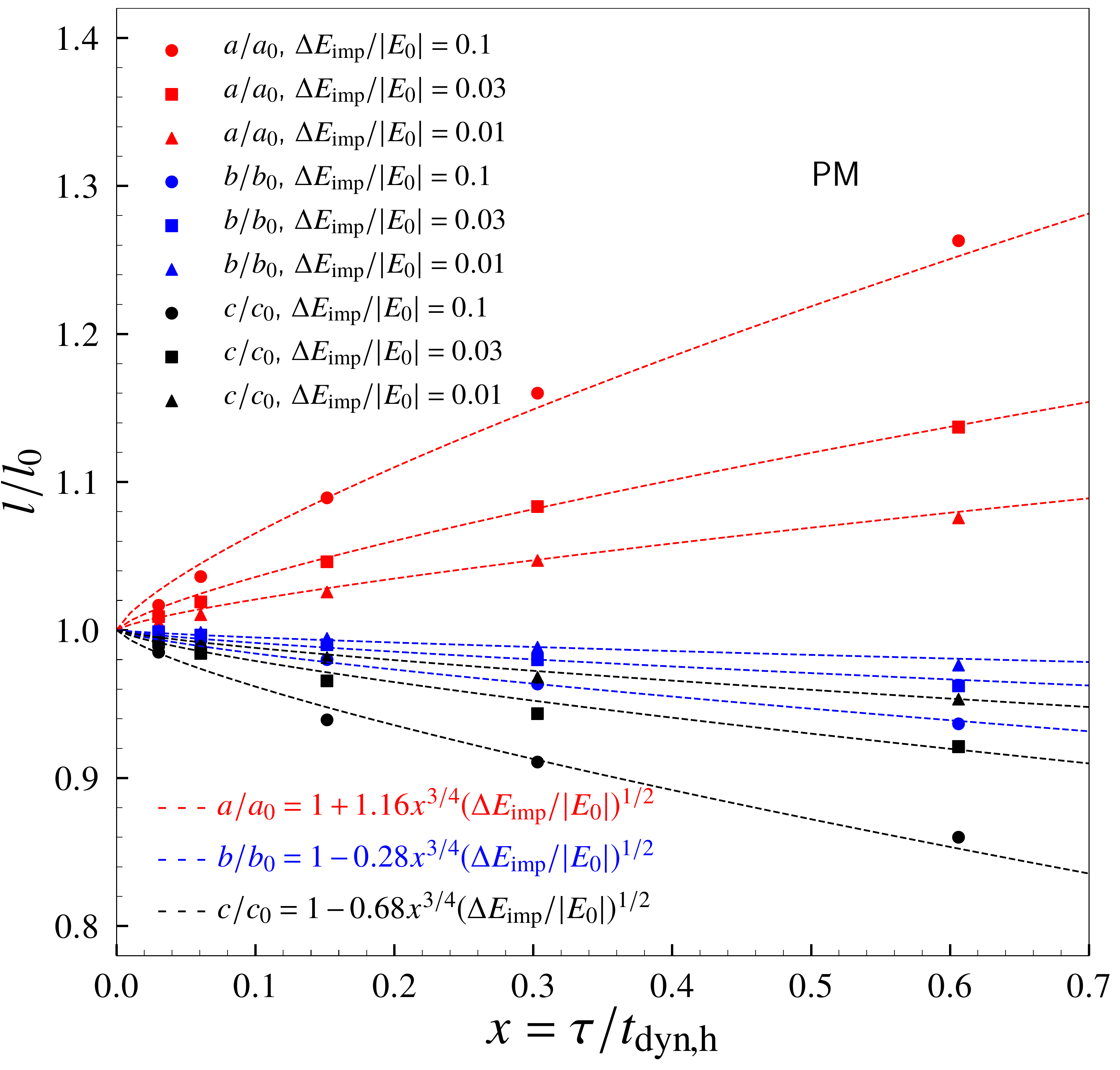}
\caption{Expected change in cluster shape for a PM shock. Symbols are the lengths of the semi-axes measured in our sets of simulations. Their values scale with the duration of the shock as $x^{3/4}$, and with the shock amplitude as $(\dEimp/|E_0|)^{1/2}$. Dotted lines are the best fits to the data using \autoref{eq:fit_shape}, which combines these two dependencies.}
\label{fig:expected_shape}
\end{figure}

\autoref{fig:expected_shape} shows the fractional change of the length of the semi-axes as a function of $x$ and different values of $\dEimp$ for a PM shock. We notice that such fractional changes scale as a power law of $\tau/\tdynh$ for a fixed shock amplitude, and as another power law of $\dEimp/|E_0|$ for a fixed duration of the shock. The power indices are similar for all axes, and average close to 3/4 and 1/2, respectively. The scaling with the energy change is expected because $\Delta v \propto \dEimp^{1/2}$ and we analyse all snapshots at the same time, such that the amount of extension/compression is proportional to $\Delta v$. The dependence of $x^{3/4}$ appears to work very well, but we do not have a physical backing for it and it resulted from trying various functional forms. The combination of these two dependencies allows us to predict the shape of the cluster after the shock as follows:
\begin{equation}
\label{eq:fit_shape}
    \frac{l}{l_0} = 1 + \mu_l \, x^{3/4} \, (\dEimp/|E_0|)^{1/2}    
\end{equation}
where $l=\{a,b,c\}$ denotes the three semi-axes. We find the values of constants $\mu_a$, $\mu_b$, and $\mu_c$ by a linear regression fit to the simulation results. This gives the best-fit values $\mu_l=\{1.16,-0.28,-0.68\}$. \autoref{fig:expected_shape} shows how well \autoref{eq:fit_shape} predicts the change in cluster shape for different values of the adiabatic parameter and shock amplitude.

It is worth noting that for a PM shock the semi-major axis always grows, while the other two decrease in size. This is because the tides exerted by a PM are extensive along the line that connects the cluster with the PM ($x$-axis), and compressive in the $z$ direction. For fast shocks there is no net tidal action in the $y$ direction \citep{spitzer58}, as seen in the left of \autoref{fig:shapePM}. The slight compression in $y$ seen in the right panel is probably because the perturbation is slower, breaking the symmetry because the tides are compressive in the first half and the cluster has time to respond. The nature of the tides is reflected in the sign of the constants $\mu_l$, and well illustrated by \autoref{fig:expected_shape}, with $a/a_0>1$, and $b/b_0,c/c_0<1$.

\begin{figure}
\includegraphics[width=\columnwidth]{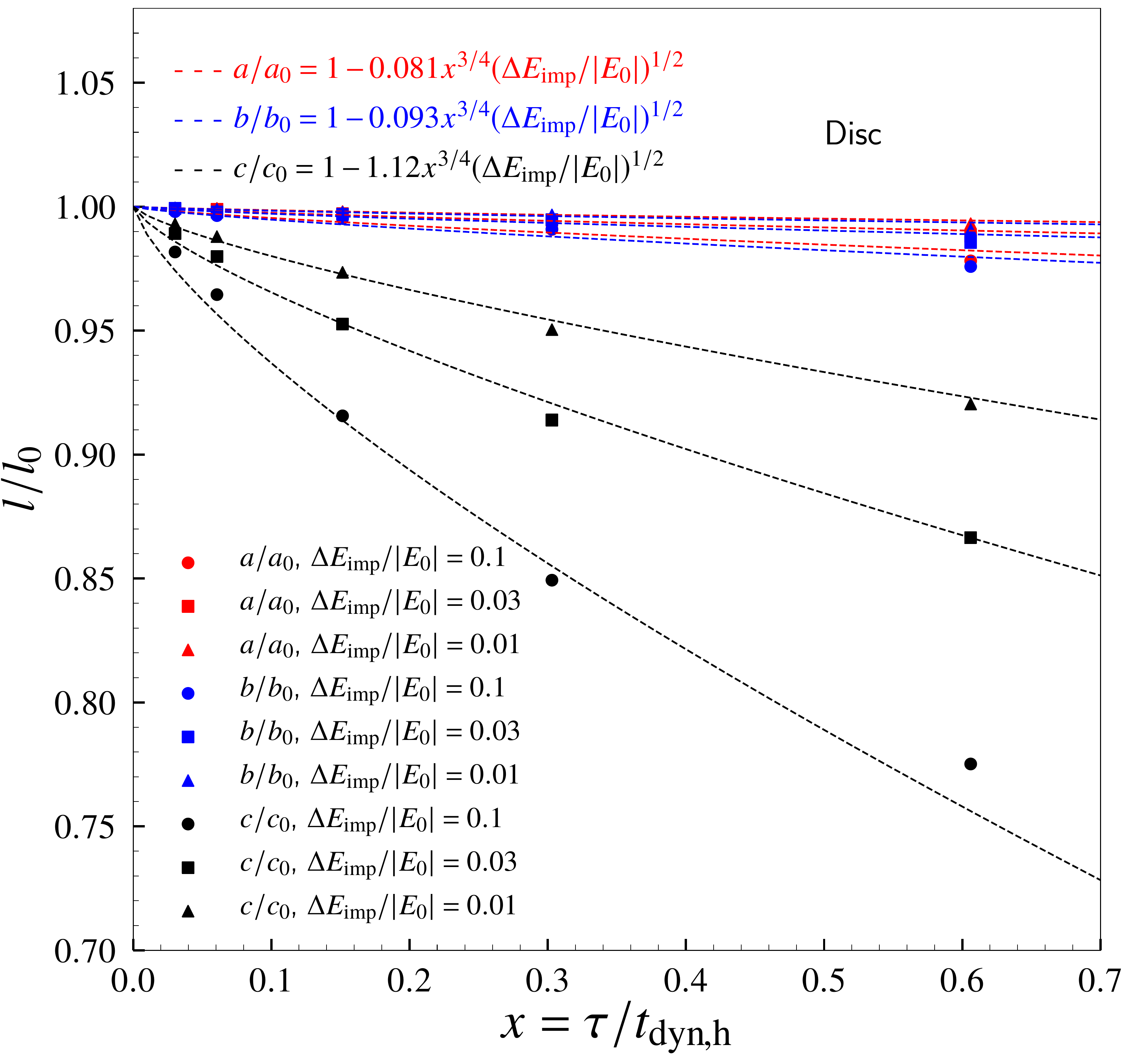}
\caption{Same as in \autoref{fig:expected_shape} but for a disc shock.}
\label{fig:expected_shape_D}
\end{figure}

We follow the same procedure for the disc case, putting together our $W_0=4$ models with shock amplitudes 0.1, 0.03, and 0.01. \autoref{fig:expected_shape_D} shows the length of the cluster semi-axes as a function of shock duration and shock amplitude. Notice that in all models the semi-lengths $a$ and $b$ change very little compared with the change in the short axis $c$. Also, the change of all three semi-lengths is negative, i.e., the cluster is compressed along its three principal axes after the shock: $a/a_0,b/b_0,c/c_0<1$.
The expected shape of the cluster after a disc shock can also be expressed by \autoref{eq:fit_shape} but with different constants $\mu_l=\{-0.081, -0.093, -1.12\}$. These three constants are negative, and reflect a larger compression in the z direction, as expected for a disc crossing.

Interestingly, the change in cluster shape described above is independent of the values of $W_0$ explored in this work. The reason for this is probably because our experiments were set up such that for a certain $\dEimp/|E_0|$, stars at a distance $r = \sqrt{\msqr}$ from the cluster centre receive the same velocity kick in clusters with different $W_0$. Because we then measure the shape parameter at that radius, it is expected to be similar across clusters with different $W_0$. The deformation at larger radii for clusters with larger $W_0$ is higher. Hence, either for a PM or disc shock, the expected shape of the cluster follows the same functional form (\autorefp{eq:fit_shape}), where the sign of the constants $\mu_l$ reflects the nature of the tides along each direction.

\subsection{Comparison with other $N$-body results}

A recent study by \citet{webb_etal19} investigated the mass loss of a star cluster experiencing a sequence of two tidal shocks separated by a time interval. The authors use the same version of \nbodytt\ code to model clusters of 50,000 stars with the Plummer density profile and two King models. They consider fast shocks and low-density clusters, such that the encounters are mostly in the impulsive regime ($x < 0.25$). The clusters are subject to shocks with only one non-zero component of the tidal tensor $T_{xx} > 0$ that varies as a step function in time, i.e. a one-dimensional extensive shock. We note that the associated mass density of this tensor is negative and that realistic perturbers have several non-zero components of the tidal tensor while the trace of the tensor is smaller than zero (for extended mass distributions) or equal to zero (for a point mass).

\citet{webb_etal19} measure the changes in cluster energy and mass 10 crossing times (40~Myr) after the shock. They find an almost exactly linear proportionality between $\Delta M$ and $\Delta E$ but with a higher normalization than our \autoref{eq:dm}: $|\Delta M|/M_0 \approx 0.4\, \Delta E/|E_0|$. The larger fractional mass loss is likely because of their choice of the tidal tensor: applying the tidal forces along a single axis results in a larger velocity increase of individual stars than when the tides act  along multiple axes (for a given energy increase, see the discussion in \autoref{ssec:dm}). This is why we find a higher fractional mass loss for the disc case than for the PM case. Because their tides are extensive, particles are pushed away from the cluster, which makes it easier to become unbound compared to our disc case, where particles still need to travel through the cluster. Another difference between our studies is in the cluster density profile: the Plummer model is different from the King models. However, \citet{gieles06} applied point-mass perturbations to Plummer models and also found $|\Delta M|/M_0 \approx 0.22\, \Delta E/|E_0|$. We think that the  larger fractional mass loss of Webb et al. is because of their tidal tensor. In addition, for such extensive tides the cluster may not have reached full dynamical equilibrium even after 10 crossing times and the value of $\Delta E$ may still be evolving.

More surprisingly, \citet{webb_etal19} find\footnote{Note that \citet{webb_etal19} performed the fit with $\Itid$ defined as the square root of our expression, and we have rescaled their slope for consistency with our definition.} a significantly non-linear relation $\Delta M/M_0 \propto \Itid^{\gamma}$ with a shallow slope $\gamma\approx0.6$. In contrast, our results in \autoref{eq:dm_itid} point to a steeper slope $\gamma>1$ for all but the most concentrated model. They attribute the non-linearity of this relation to a substantial escape time for newly unbound stars. While the escape time effect plays a role in decreasing mass loss in tidally limited clusters \citep[e.g.,][]{lee_ostriker87, baumgardt01}, it is less likely to affect isolated clusters studied by \citet{webb_etal19}. We therefore conclude that the escape time is not the main cause of the non-linearity of $\Delta M(\Itid)$. Note that in our results the $\Delta M(\Delta E)$ relation is as non-linear as $\Delta M(\Itid)$, while $\Delta E$ is exactly linearly proportional to $\Itid$ as expected from the tidal theory. 

We also note that in this work we considered a spherical perturber as infinitely compact PM, appropriate for dense molecular clouds or star-forming regions. More extended perturbers, such as a galactic bulge or a satellite galaxy, may lead to more significant adiabatic damping of the energy change \citep[c.f.][]{gnedin_etal99b}.

\section{Summary}
\label{sec:summary}

We studied the response of a star cluster to different tidal perturbations by using direct $N$-body simulations. We measured the energy gain of the cluster as it experiences a PM flyby or a disc crossing. We used different density profiles given by \citet{king66} models with dimensionless central concentration parameters $W_0 = [2, 4, 6, 8]$. Also, we explored the impulsive and adiabatic regimes by varying the duration of the encounter.

For fast shocks the predictions from the impulse approximation are recovered for both perturber types and all cluster concentrations. However, in the adiabatic regime we find important differences.

For slow disc crossings, we find a smaller energy increase than in the impulsive regime. This has previously been attributed to the effect of adiabatic damping \citep{gnedin_ostriker99}. We show that the energy change depends on the cluster's concentration (i.e. $W_0$), in a way that is consistent with adiabatic damping: clusters with less stars in the impulsive regime have a smaller energy increase. However, this $W_0$ dependence in the energy change can also be attributed to a geometrical distortion (i.e. compression) in the first half of the perturbation, which reduces the importance of the remaining part of the perturbation because the cluster is compressed along the $z$ direction. For similar reasons, slow PM perturbations lead to much larger energy increases, comparable or even larger than what is expected in the impulsive regime. Because the PM tidal forces have an extensive component, this result can be explained by geometrical distortion. Since we see a similar correlation of $\Delta E$ with $W_0$, we conclude that adiabatic damping does play a role, of similar importance to the geometrical distortion.

We present an accurate parameterization of the expected changes in cluster shape as a function of the amplitude of the perturbation and the duration of the shock (\autorefp{eq:fit_shape}).

These results are useful for analytical modeling of cluster disruption for a sub-grid models of cluster evolution in numerical simulations of galaxy formation. We conclude that the following steps are necessary to correctly model the impact of tidal shocks. To compute the energy gain $\Delta E$, it is important first to identify whether the perturbation is compressive (disc-like) or has an extensive component (PM-like). For a disc-like perturbation, $\Delta E$ is computed taking into account the cluster density profile as expressed in \autoref{eq:ad_corr2} and illustrated in \autoref{fig:Fit_3_2}. On the other hand, for a PM-like perturbation the value of $\Delta E$ is nearly constant for $x<0.7$, then it increases as a function of $x$ and $W_0$.

To understand the mass evolution, the most important assumption is the cluster density profile, which affects the resulting mass loss (for a given $\Itid$) by more than an order of magnitude. Semi-analytic models of cluster evolution with tidal perturbations usually assume a fixed value for the $W_0$ parameter \citep[e.g.][]{gieles06, 2010ApJ...712L.184E, pfeffer_etal18}. To increase the predictive power of such models, we need better understanding of the evolution of the density profile of clusters evolving in a Galactic tidal field, experiencing repeated shocks, two-body relaxation and stellar evolution.

\section*{Acknowledgements}
We thank Jeremy Webb for helpful comments on the manuscript, and the referee, Douglas Heggie, for a careful review and insightful suggestions.
LAMM acknowledges support from PAPIIT IA101520 grant. MG and LAMM are acknowledge support from the European Research Council (ERC StG-335936, CLUSTERS). MG acknowledges support from the Ministry of Science and Innovation through a Europa Excelencia grant (EUR2020-112157). OG is supported in part by NSF through grant 1909063. HL is supported by NASA through the NASA Hubble Fellowship grant HST-HF2-51438.001-A awarded by the Space Telescope Science Institute, which is operated by the Association of Universities for Research in Astronomy, Incorporated, under NASA contract NAS5-26555.

\section*{Data availability}

The data that support the findings of this study are available from the corresponding author, upon reasonable request.

\bibliographystyle{mnras}
\bibliography{gc,tidal} 


\bsp	
\label{lastpage}
\end{document}